\documentclass[a4paper,12pt]{article}
\usepackage{latexsym}
\usepackage{amsmath,amssymb,bm,ascmac,color,calc}
\usepackage[mathscr]{eucal}
\usepackage{graphicx,color}
\usepackage{cite}
\usepackage{lineno}
\usepackage{setspace}
\usepackage{etoolbox}
\usepackage{url}
\usepackage{longtable} 
\usepackage{enumerate}
\usepackage{multirow,booktabs}
\usepackage[affil-it]{authblk}
\usepackage[normalem]{ulem} 

\allowdisplaybreaks 

\begin{document}

\begin{titlepage}

\title{{\Large\bf The potential of the ILC beam dump for high-intensity and large-area irradiation field with atmospheric-like neutrons and muons}\vspace{1.2em}}

\author[]{\normalsize Yasuhito~Sakaki}
\author[]{\normalsize Shinichiro~Michizono}
\author[]{\normalsize Nobuhiro~Terunuma}
\author[]{\normalsize Toshiya~Sanami}

\affil[]{\small High Energy Accelerator Research Organization (KEK), Ibaraki 305-0801, Japan.}

\date{}

\maketitle

\vspace{1cm}

\begin{abstract}
We evaluate the neutron and muon fluxes produced in the ILC beam dumps by Monte Carlo simulations and discuss their potential use in irradiation fields. We show that the beam dumps can provide high-intensity neutron and muon fluxes with spectra quite similar to secondary cosmic rays, which are suitable for soft error studies. The beam dumps deliver neutrons about $10^{11}$ times the secondary cosmic rays on spaces perpendicular to the beam axis and muons $10^8$ times downstream of the beam dumps in the initial phase of the ILC. Large-area irradiation of 1\,m$^2$ or more is possible. Differences in the energy distribution of muons in electron and positron beam dumps are also discussed for particle physics experiments.
\end{abstract}

\thispagestyle{empty} 

\end{titlepage}

\hrule
\tableofcontents
\vskip .2in
\hrule
\vskip .4in

\setlength{\parskip}{6.5pt}%

\section{Introduction}
The International Linear Collider (ILC)~\cite{Behnke:2013xla, Baer:2013cma, Adolphsen:2013jya, Adolphsen:2013kya, Behnke:2013lya} is one of the proposed next-generation electron-positron colliders. The main goal of the ILC is to study events of electron-positron collision to perform high-precision measurements of the Higgs bosons and search for new particles produced by the electroweak interaction. In recent years, there has been active discussion on using ILC infrastructures for applications other than the primary purpose~\cite{ILCX2021, ILCInternationalDevelopmentTeam:2022izu}. The facility that has received the most attention in the discussions is the beam dumps that absorb the accelerated beam. These are installed at the end of electron and positron beamlines.

The use of the ILC beam dumps has excellent potential.\footnote{About 15 beam dumps will be installed at the ILC. The ones in focus in the paper are called the ``main beam dumps". In this paper, they are referred to simply as ``beam dumps".} The main feature of the ILC beam dumps is that the beam is constantly fed into them during operation because the ILC is a linear collider. At an initial stage of the ILC, 125~GeV electron or positron beams are delivered to the beam dump at 2.6~MW. When a high-energy electron-positron beam is injected into the beam dump, many bremsstrahlung photons are produced in the electromagnetic shower, which induce various secondary particles. They can be exploited for particle physics, nuclear physics, condensed matter physics, and industrial applications.

Another feature is that the ILC beam dump is a water cylinder about 10~m long. The standard atmospheric pressure at sea level is 1013.25~hPa, so the equivalent of 10.3~m of water, which is approximately the same height as the ILC beam dump, is piled up on the ground. Therefore the ILC beam dump can be regarded as a secondary particle generator that compresses 100~km of the Earth's atmosphere to about 10~m, and the produced particle spectra are expected to be atmospheric-like. Although the difference is that secondary cosmic rays are due to hadronic interaction and the particle field around the ILC beam dump is due to electromagnetic interaction, comparing the spectrum of particles produced in the ILC beam dumps with that of secondary cosmic rays is interesting.

The main purpose of this paper is to evaluate the potential of the ILC beam dumps as an irradiation field, especially for soft error studies. We evaluate the fluxes of neutrons and muons produced in the beam dumps. Accelerator-driven beams of these particles help study soft error in semiconductor systems. The soft error is a temporary malfunction of semiconductors, mainly caused by cosmic ray neutrons, which have recently become more common as the amount of charge held by memory has decreased with the miniaturization of semiconductors~\cite{556861, 903813}. There is also growing concern that cosmic ray muons may be a source of the soft error on the ground~\cite{hubert2015impact, infantino2016monte}.

The ILC beam dump has the potential to provide a distinctive irradiation facility that is capable of delivering high intensities, large irradiation areas, atmospheric spectra and multiple particle species simultaneously, which is not possible with existing irradiation facilities. It would help test the soft error for integrated systems, such as self-driving cars, that require a high level of safety. Existing neutron beam facilities~\cite{doi:10.13182/NSE90-A27471, NOWICKI2017374, blackmore2003improved, RCNP_white_neutron, andreani2008facility} struggle to obtain atmospheric spectra. In addition, no muon irradiation facilities provide large irradiation areas and atmospheric spectra.

This paper also provides important input for the currently proposed beam dump experiments for particle physics at the ILC~\cite{Kanemura:2015cxa, Sakaki:2020mqb}. If the experiment's detector systems are placed around the beam dump, the neutron flux is an essential information to assess radiation damage to the detectors. In those experiments, the forward muons produced in the beam dumps provide the background and trigger signal events. Therefore, evaluating the muon flux is necessary to design muon shieldings and estimate the physics sensitivity. Also, one of the features of the ILC is the ability to use high-energy positron beams, which pair annihilate with atomic electrons in the beam dumps, causing muon pair production. The contribution of this process to the whole is worth investigating.

This paper is organized as follows. In Section~2, ILC beam dumps and beam conditions are described. In Section~3, we present our simulation setup, particularly the simulation geometries. In Section~4, details of numerical results are shown. Section~5 is devoted to the summary.

\section{ILC beam dump and beam conditions}
\label{sec:ILC}

The base design of the current ILC beam dump is a cylindrical tank, 1.8~m in diameter and 11~m long, filled with water as an absorber~\cite{Satyamurthy:2012zz}. The thickness of the side of the tank is 1~inch, and the thickness of the bottom is 3~inches. There are three pipes inside the tank, two inlet and one outlet. The inlet pipes have an inner diameter of 8~inches and have a fin structure to blow out the cooling water. The outlet pipe has an inner diameter of 10~inch and have numerous pores. Cooling water is collected through the pores in the outlet pipe. 

Table~\ref{table:ILC} shows some beam conditions expected in the ILC. 
Starting from the ``Initial" stage, various upgrades are being considered. In the table, $\mathcal{L}$-Upgrades refer to luminosity upgrades, which increase the number of particles transported; in the $Z$-pole stage, energy is reduced to maximize the frequency of $Z$ boson production. Further upgrades are planned to increase energy and luminosity further.
It shows that the ILC beam dumps absorb beam powers in the order of MW. Therefore, thick shielding is required between the beam dump and the maintenance area to prevent radiation exposure. This study mainly assumes beam conditions at the initial stage, i.e. 125 GeV and 2.6 MW. We also show results for different beam energies of 45.6 and 500 GeV.

\begin{table}[!t]
 \centering
 \caption{Beam energy and average beam power injected into each beam dump for the 125 GeV initial stage configuration and possible upgrades.}
 \vspace{7pt}
 \label{table:ILC}
 \small
 \begin{tabular}{c||c|c|c|ccc}
  Quantity & Initial & $\mathcal{L}$-Upgrade & $Z$-pole & & Upgrades &  \\
  \hline
  Beam energy [GeV] 
  	& 125 & 125 & 45.6 & 250 & 125 & 500 \\
	\begin{tabular}{c}
	Average beam power [MW]\\(per line)
	\end{tabular}
   	& 2.6 & 5.2 & 0.7, 1.4 & 5.2, 10.5 & 10.5 & 13.6 \\
 \end{tabular} 
\end{table}

\section{Simulation setup}

\subsection{Geometries for simulation}

\begin{figure}[!t]
\begin{center}
\includegraphics[width=7.5cm, bb=0 0 465 253]{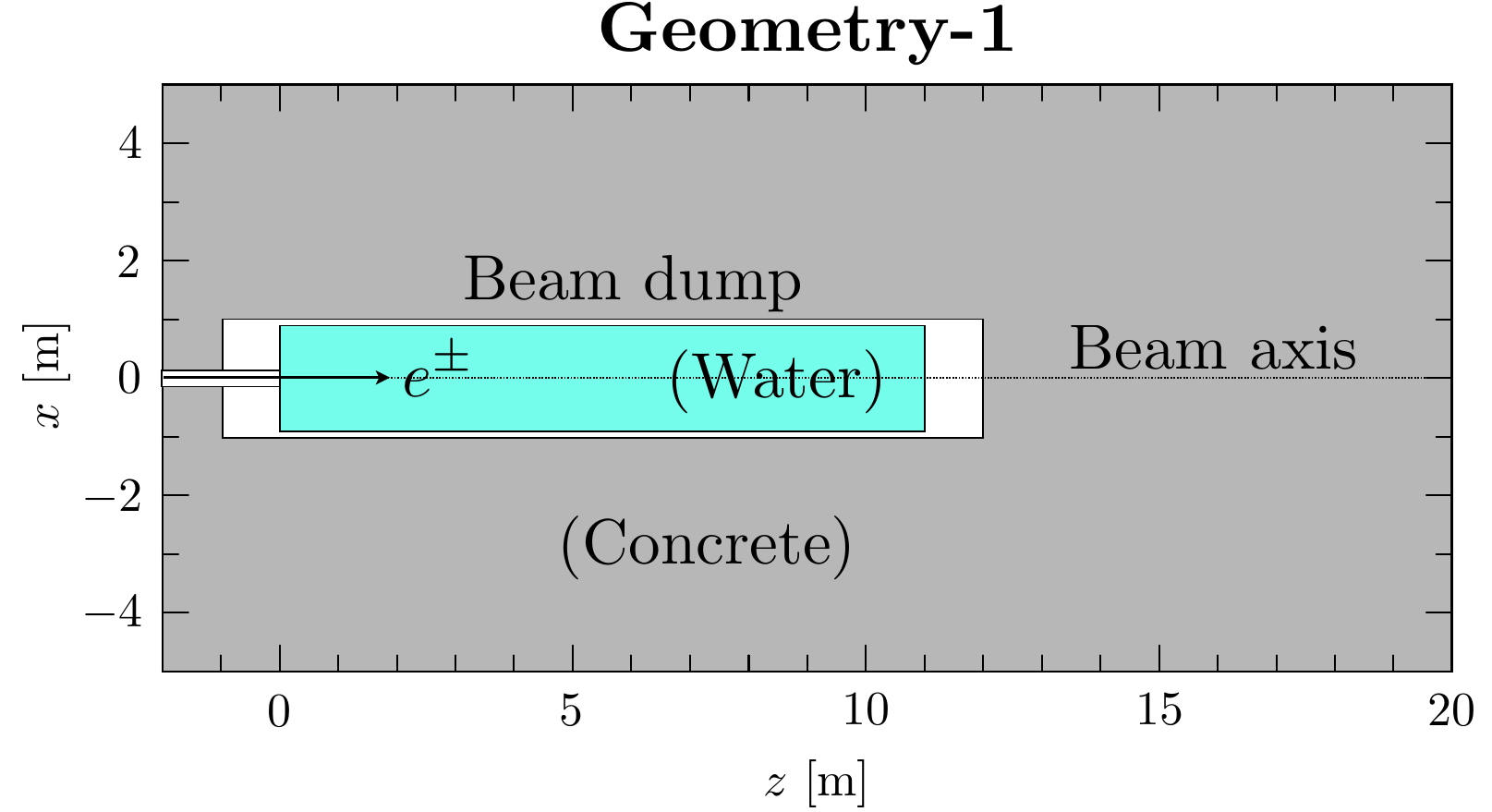}\\
\includegraphics[width=7.5cm, bb=0 0 465 253]{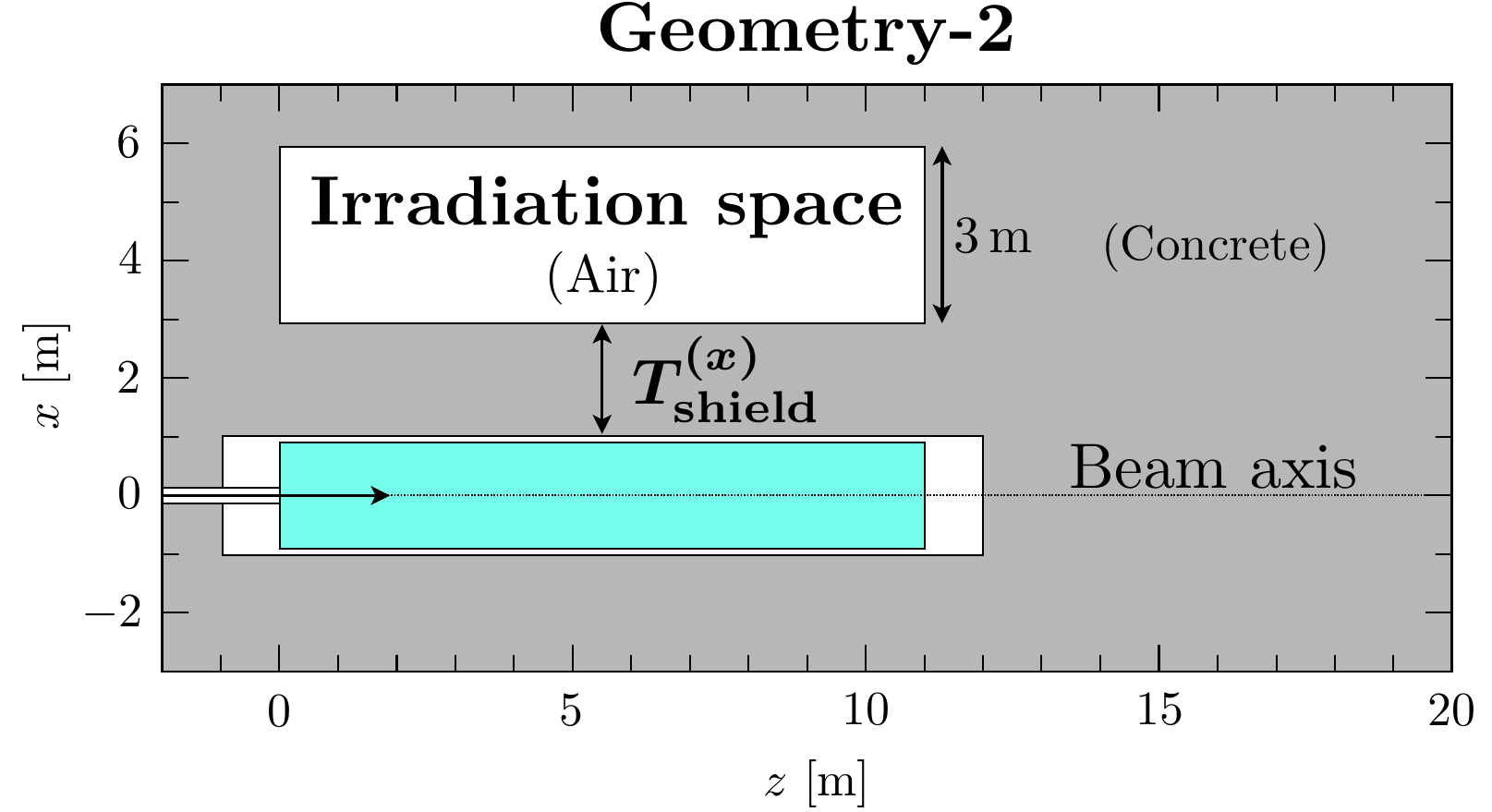}
\includegraphics[width=7.5cm, bb=0 0 465 253]{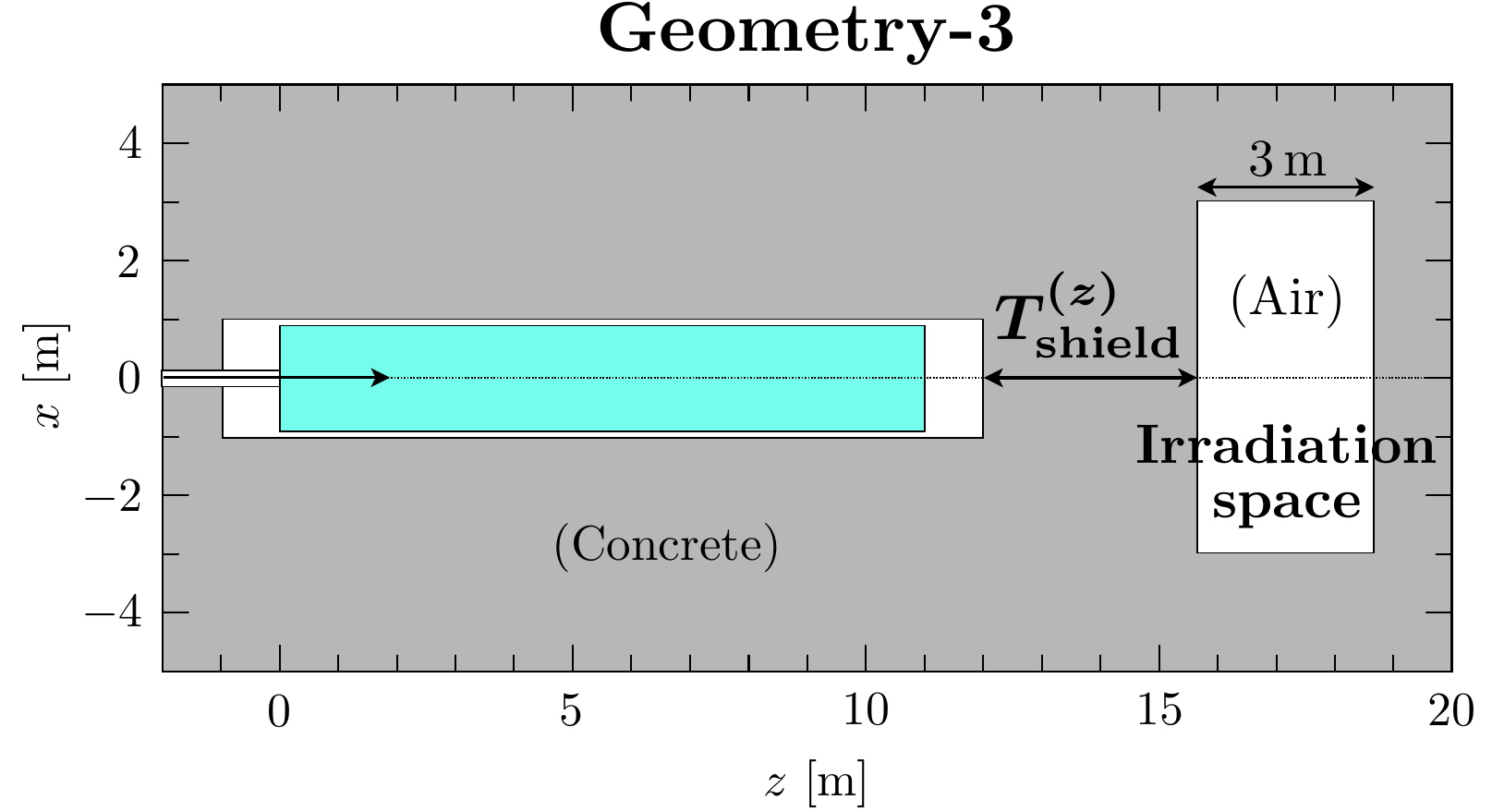}
\caption{Three geometries used in flux evaluation. Geometry-2 and -3 have parameters of concrete shield thickness $T_{\rm shield}^{(x)}$ and $T_{\rm shield}^{(z)}$ and irradiation spaces.}
\label{Geometry}
\end{center}
\end{figure}

It is possible to provide irradiation spaces at a location perpendicular to the beam axis or downstream of the beam dump. We perform simulations in the following three geometries to examine the neutron and muon fluxes:
\begin{enumerate}[{\bf Geometry-1}]
 \item Uniform concrete shield around beam dump;
 \item Concrete shield of thickness $T_{\rm shield}^{(x)}$ perpendicular to beam axis, and irradiation space placed behind it;
\item Concrete shield of thickness $T_{\rm shield}^{(z)}$ downstream of beam dump, and irradiation space placed behind it.
\end{enumerate}
The three geometries are described in Fig.~\ref{Geometry}.

Calculations using Geometry-1 provide an overall characterization of the neutron and muon fluxes around the dump and their attenuation. This will help to determine the optimum location of the irradiation field. For simplicity, the shield is uniformly made of concrete. Although, in reality, iron shields are likely made of different parts, their length can be converted into an equivalent length of concrete. For muons, the stopping power of steel is about 3.2~times higher than that of concrete, which results in a shorter attenuation distance. For neutrons, iron has an interaction length approximately 2.5~times shorter than that of concrete and therefore has a shorter attenuation distance.\footnote{The density of the concrete used in the calculations is 2.3~g/cm$^3$, and the composition is standard~\cite{RAD1}.}

Geometry-2 and -3 have a space with a width of 3~m, assuming irradiation of large volumes such as cars. The results can be inferred to some extent from the results of Geometry-1. However, the flux spread and energy spectrum are slightly different in the concrete and the space. For this reason, calculations are also carried out for Geometry-2 and -3, which are close to the expected shape of the irradiation field.

\subsection{Monte Carlo tool}
The {\tt PHITS} code~\cite{Sato:2018} is used in this study. Electromagnetic showers make use of {\tt EGS5}~\cite{Hirayama:2005zm}, which is implemented in {\tt PHITS}. Neutrons are mainly produced by photons in electromagnetic showers or secondary hadrons interacting with nuclei or nucleons. In {\tt PHITS}, they are calculated based on {\tt JAM}~\cite{Nara:1999dz}, {\tt JQMD}~\cite{PhysRevC.52.2620}, and {\tt GEM}~\cite{FURIHATA2000251} models modified for photonuclear reactions. {\tt INCL4.6}~\cite{PhysRevC.87.014606} is used for the transport of hadrons such as neutrons. These models have been validated with various experimental results~\cite{iwamoto2017benchmark, matsuda2011benchmarking}. 
For the interaction of neutrons with nuclei below 200~MeV, an Evaluated Nuclear Data Library, {\tt JENDL4.0/HE}~\cite{doi:10.1080/18811248.2011.9711675, kunieda2016overview}, is used. It is one of the best ways for neutron transport in that energy range. 

The intensity of neutrons produced in the electron beam dump and their attenuation in the shield has been comprehensively studied at SLAC~\cite{JENKINS1979265}. This study and {\tt PHITS} are widely used in the shielding design of operational electron accelerator facilities, and their consistency has been confirmed in the operation of accelerator facilities.

Muons are mainly produced by pair production from a real photon in electromagnetic showers, pair production from a positron above 43.7\,GeV, and $\pi^{\pm}/K^{\pm}$ decay. The differential cross section obtained by QED's perturbation theory is implemented in {\tt PHITS} for the former two. The calculation accuracy of the muon pair production from a real photon, which is the dominant process, has been verified with high accuracy~\cite{Sakaki:2020cux} by the data of the muon shielding experiment at SLAC~\cite{Nelson:1974tu, Nelson:1974tv}.

For neutron and muon production, the direct electron production due to virtual photons is negligible in thick targets, such as beam dumps.

\section{Numerical results}
The neutron and muon fluxes in the simulation setup of the previous section are evaluated. This section considers the energy distributions of the neutron and muon fluxes and their integrations. The energy range of neutrons is broad and extends into the thermal neutron regime ($\sim 0.01$\,eV at 300\,K). We consider neutrons above 10\,MeV. This is because neutrons in that energy range have a large soft error cross section~\cite{iwashita2020energy}, the amount of which is important for assessing the performance of the irradiation field for soft errors. 

Some residual photons from electromagnetic showers remain in regions close to the beam dump, and the photon spectrum is not atmospheric-like. However, placing appropriate metal shieldings can easily control the electromagnetic shower component. This paper does not discuss the optimization of the shielding setup and photon fluxes. In the near-surface atmosphere and shields, photons other than electromagnetic showers are mainly due to neutrons. So if the neutron spectrum is atmospheric-like, the photon flux will also be atmospheric-like.

\begin{figure}[!t]
\begin{center}
\includegraphics[width=7.5cm, bb=0 0 536 252]{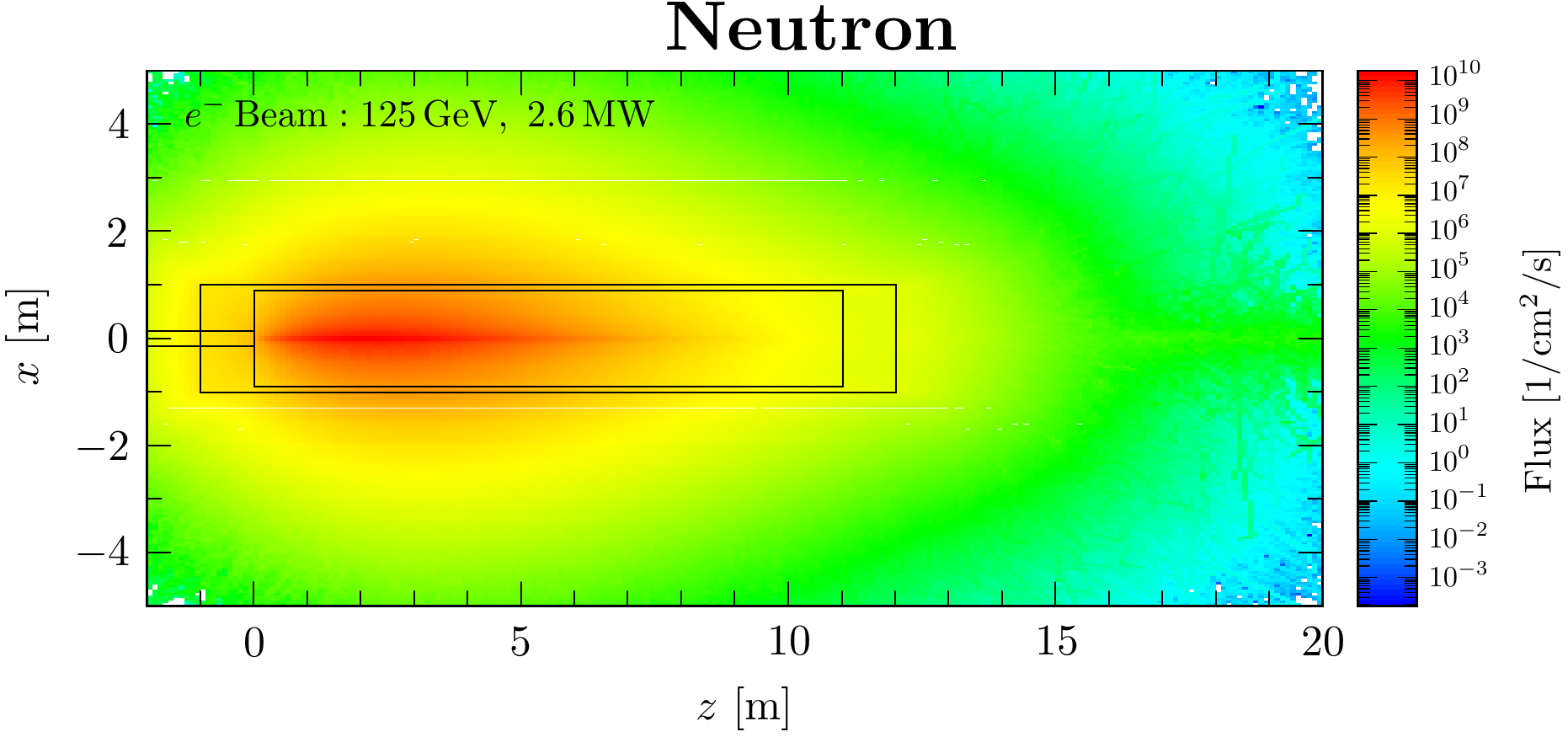}~
\includegraphics[width=7.5cm, bb=0 0 536 252]{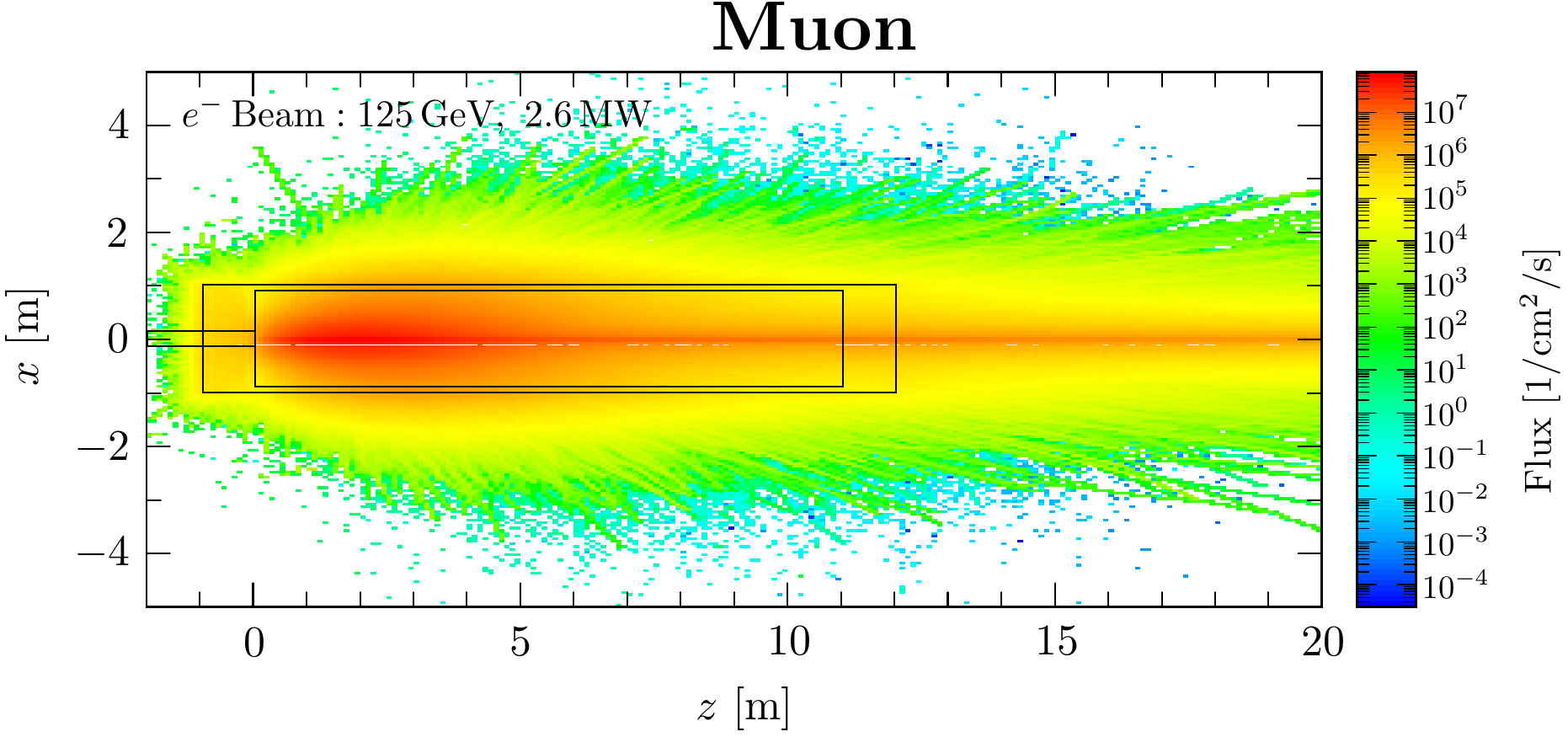}
\caption{Neutron and muon fluxes around the beam dump in Geometry-1.}
\label{2d}
\end{center}
\end{figure}

\begin{figure}[!t]
\begin{center}
\includegraphics[width=7.5cm, bb=0 0 326 283]{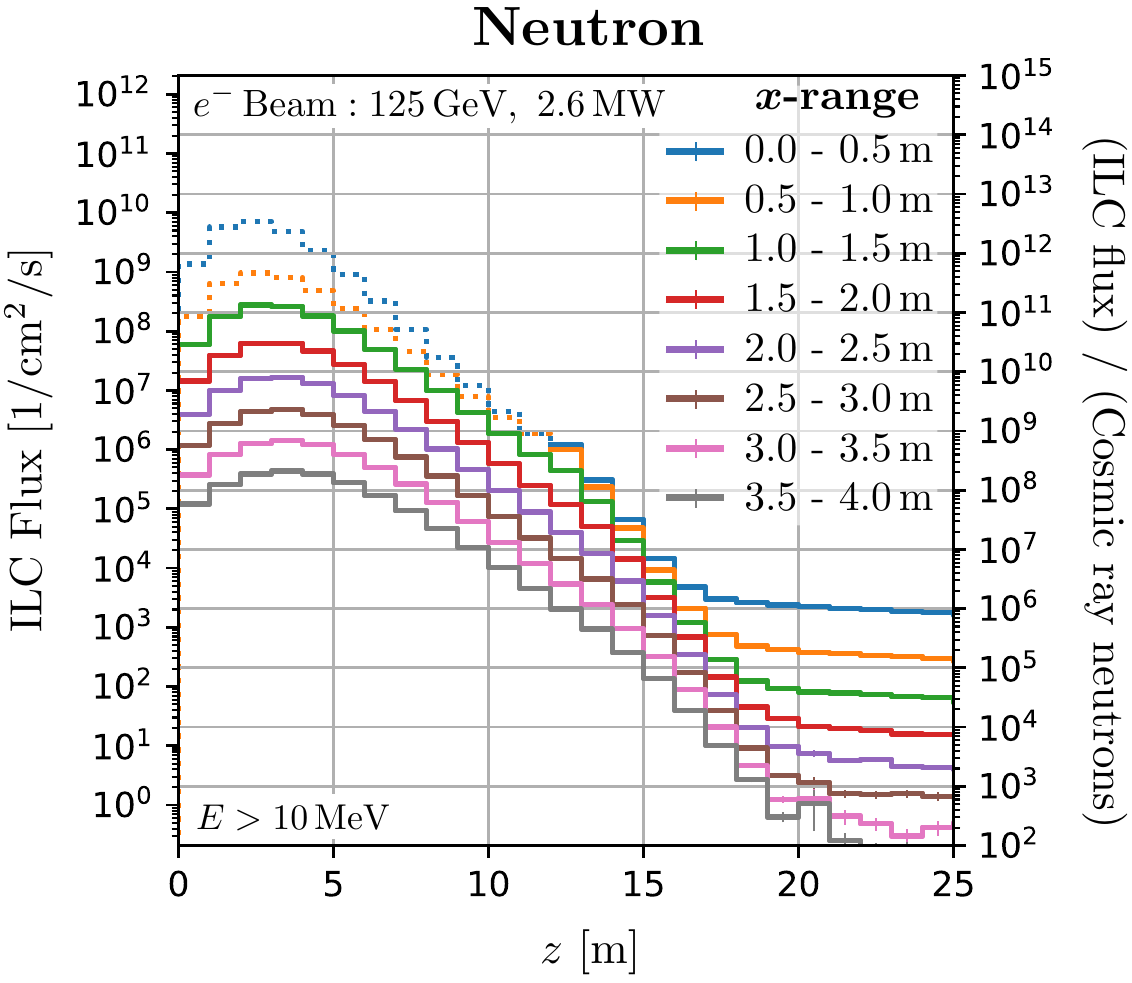}~~~
\includegraphics[width=7.5cm, bb=0 0 326 283]{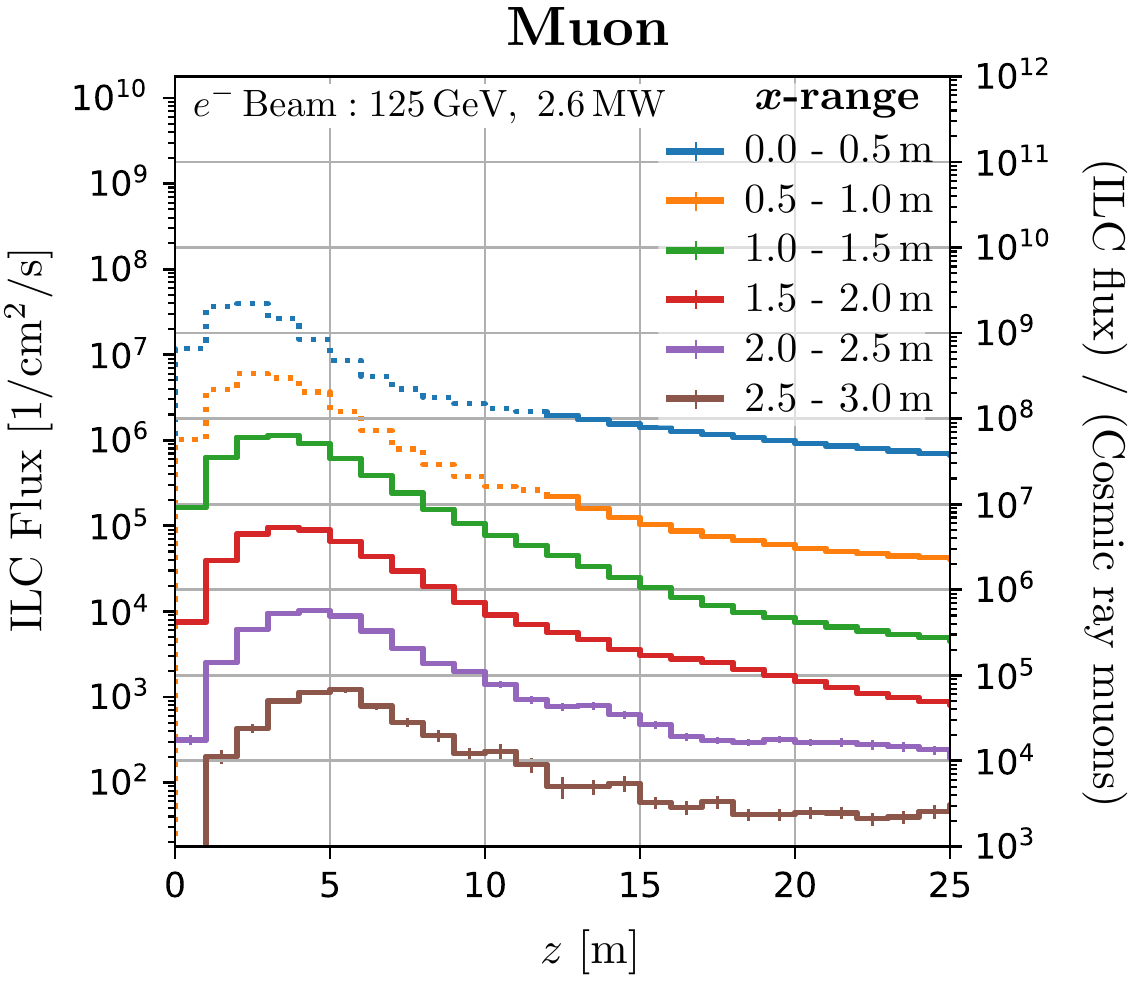}
\caption{The $z$-distribution of the neutron and muon fluxes at each $x$-range for Geometry-1. The beam energy and power are 125~GeV and 2.6~MW. The right $y$-axis shows the ratio of the flux to the secondary cosmic ray at sea level. Results in the beam dump area that cannot be used as irradiation space are represented by dotted lines.}
\label{z}
\end{center}
\end{figure}

\subsection{Geometry-1}

Figure~\ref{2d} shows the density distribution of the fluxes in Geometry-1. The beam axis is assigned to the $z$-axis, and its perpendicular direction to the $x$-, $y$-axis. The $x$ and $y$ coordinates of the beam axis are zero.

Neutrons are emitted almost isotropically from the beam dump. It can be seen from the left panel of Figure~\ref{2d} that the core of neutron production is in the region between 2~-~4~m in the beam direction from the beam injection point. Since the radiation length of water is 36~cm, that region corresponds to a length of 6~-~11~radiation length, where the photons in the evolved electromagnetic shower produce neutrons through photonuclear reactions. We, therefore, see that the high-intensity neutrons can be obtained close to the neutron emission region on the side of the beam dump.

As shown in the right panel of Figure~\ref{2d}, muons tend to be scattered in the beam direction. These forward muons are mainly produced by pair production from the bremsstrahlung. This shows the potential for a high-intensity muon irradiation field downstream of the beam dump on the beam axis.

We quantify the production and attenuation of neutrons and muons. In Figure~\ref{z}, the $z$ distribution of the flux at each $x$-range is shown. The $y$-direction is integrated in the region $-50~{\rm cm}<y<50~{\rm cm}$. The right $y$-axis of the plots shows the flux ratio to the secondary cosmic ray at sea level in Tokyo. Fluxes in the beam dump area that cannot be used for irradiation are represented by dotted lines. In the initial phase of the ILC, it can be seen that the beam dump can deliver neutrons $\sim10^{11}$ times the secondary cosmic ray on the region perpendicular to the beam axis and muons $\sim10^8$ times downstream of the beam dumps.

\subsection{Geometry-2: Flux in perpendicular space}
\label{sec:Geometry-2}

\begin{figure}[!t]
\begin{center}
\includegraphics[width=7.5cm, bb=0 0 326 283]{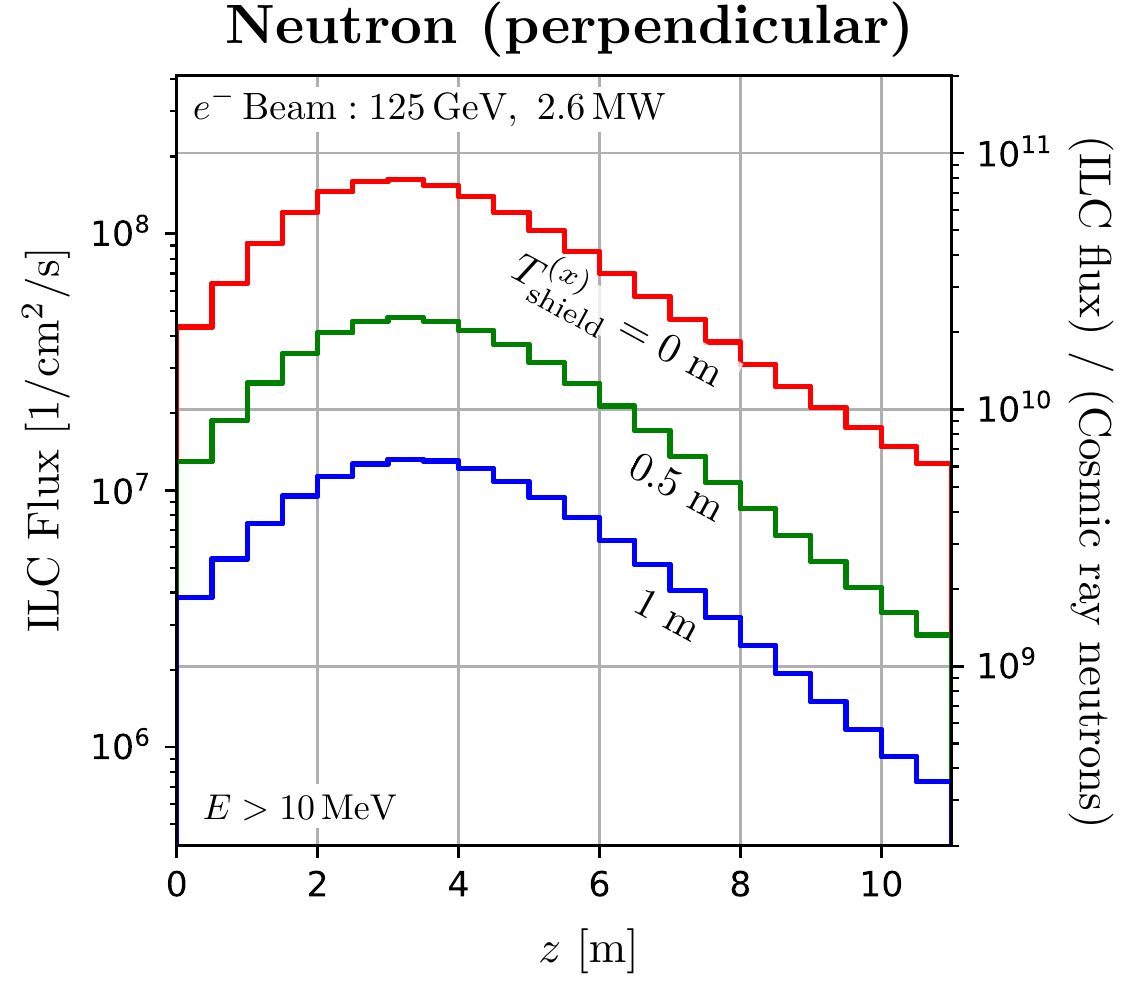}~
\includegraphics[width=7.5cm, bb=0 0 326 283]{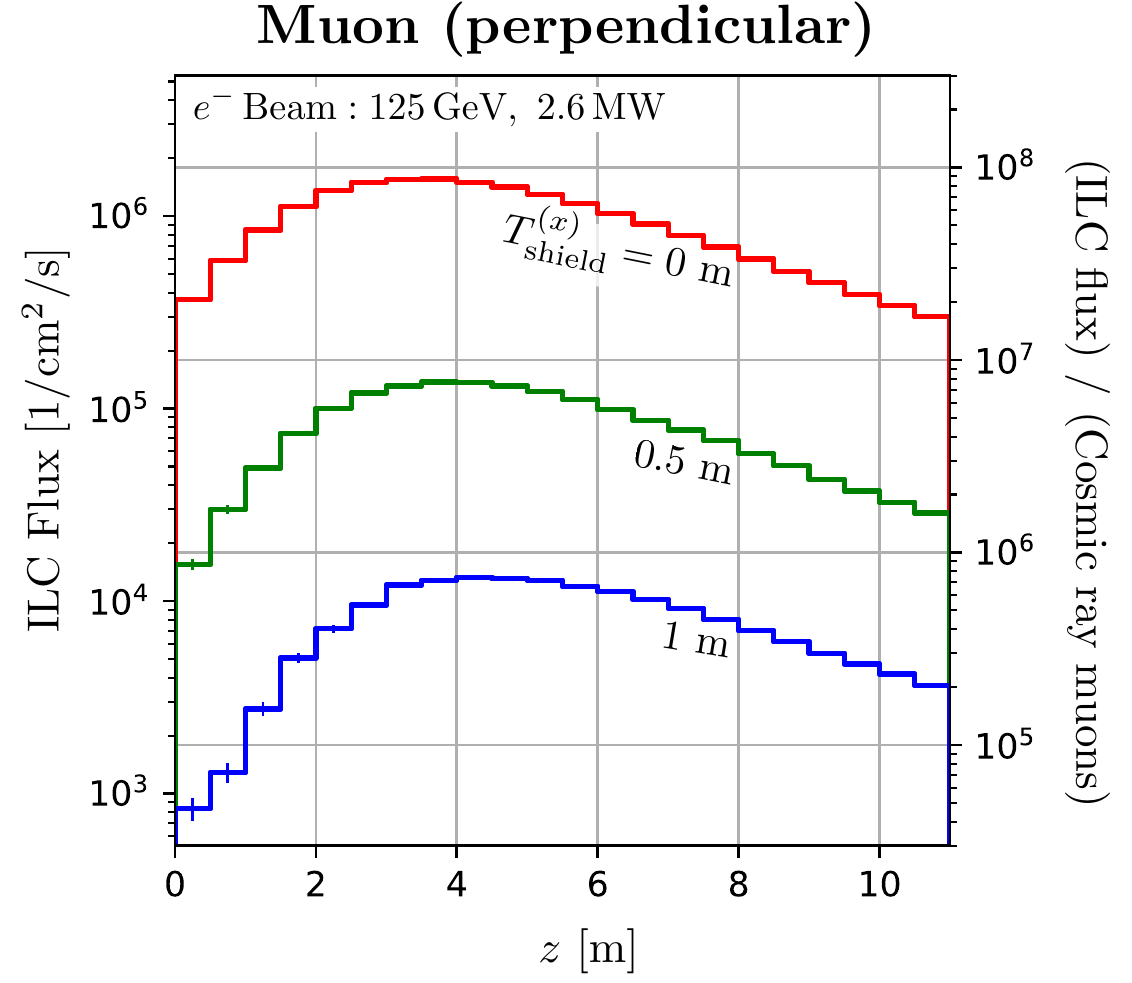}
\caption{The $z$-distribution of the fluxes in the irradiation space placed perpendicular to the beam axis using Geometry-2.}
\label{side_z}
\end{center}
\end{figure}

We perform calculations using Geometry-2 to investigate the flux and energy distribution of neutrons and muons in the irradiation space placed perpendicular to the beam axis. Figure~\ref{side_z} shows the $z$-dependence of the flux in that space. The $x$-direction is integrated over the whole space, and the $y$-direction over the region $-50~{\rm cm}<y<50~{\rm cm}$. The thickness of the concrete shield is chosen for $T_{\rm shield}^{(x)}=0, 0.5, $ and 1~m. The flux is found to be maximum at $z=3\text{ - }4$~m. The variation of neutron flux intensity is within an order of magnitude over 10~m length, which enables uniform large-area neutron irradiation.

Neutron fluxes decrease by one order of magnitude for a concrete shield of 1~m, while muons 0.5~m. The muons in the perpendicular space are mainly due to the decay of charged pions produced in photonuclear reactions. Charged pions emitted with a wide production angle have low energy due to kinematics and energy loss in materials. As a consequence, muons have low energy and penetration power. Therefore, the number of muons can be adjusted by the thickness of the shield without reducing neutrons. 

\begin{figure}[!t]
\begin{center}
\includegraphics[width=7.0cm, bb=0 0 283 283]{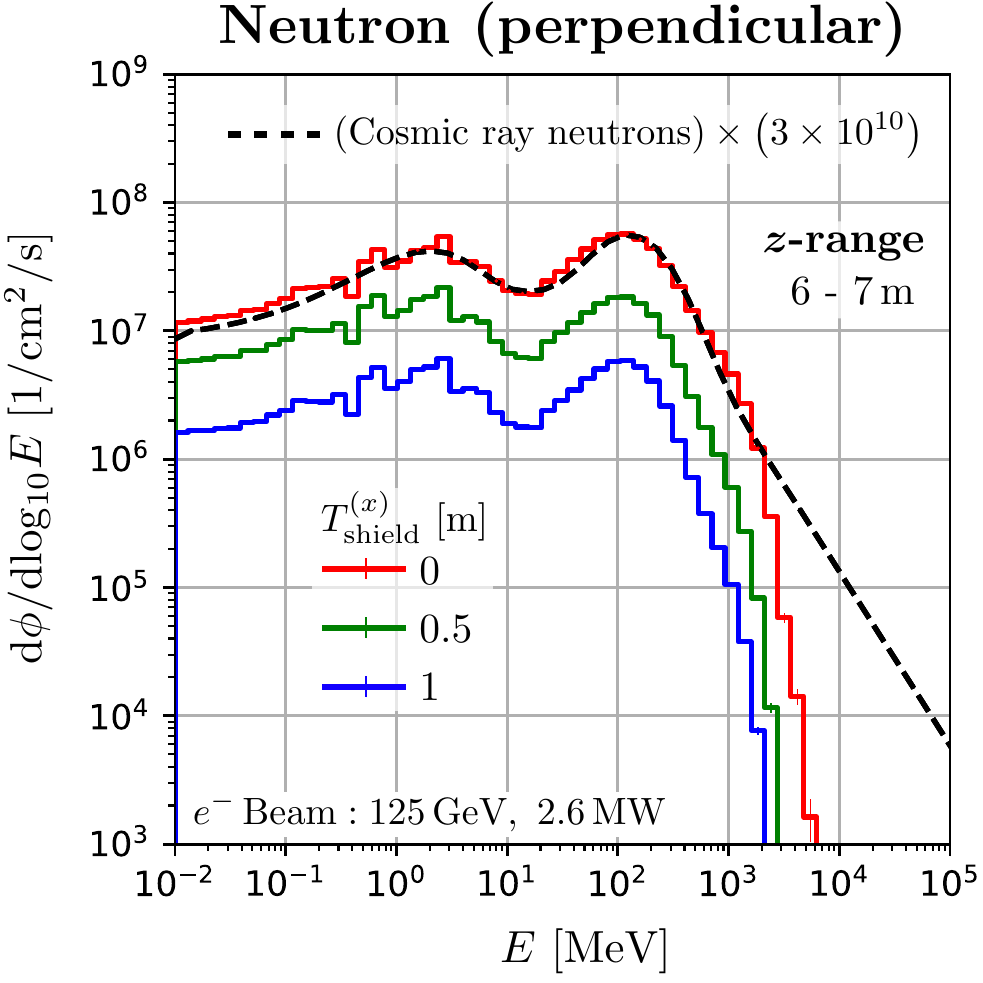}~~~
\includegraphics[width=7.0cm, bb=0 0 283 283]{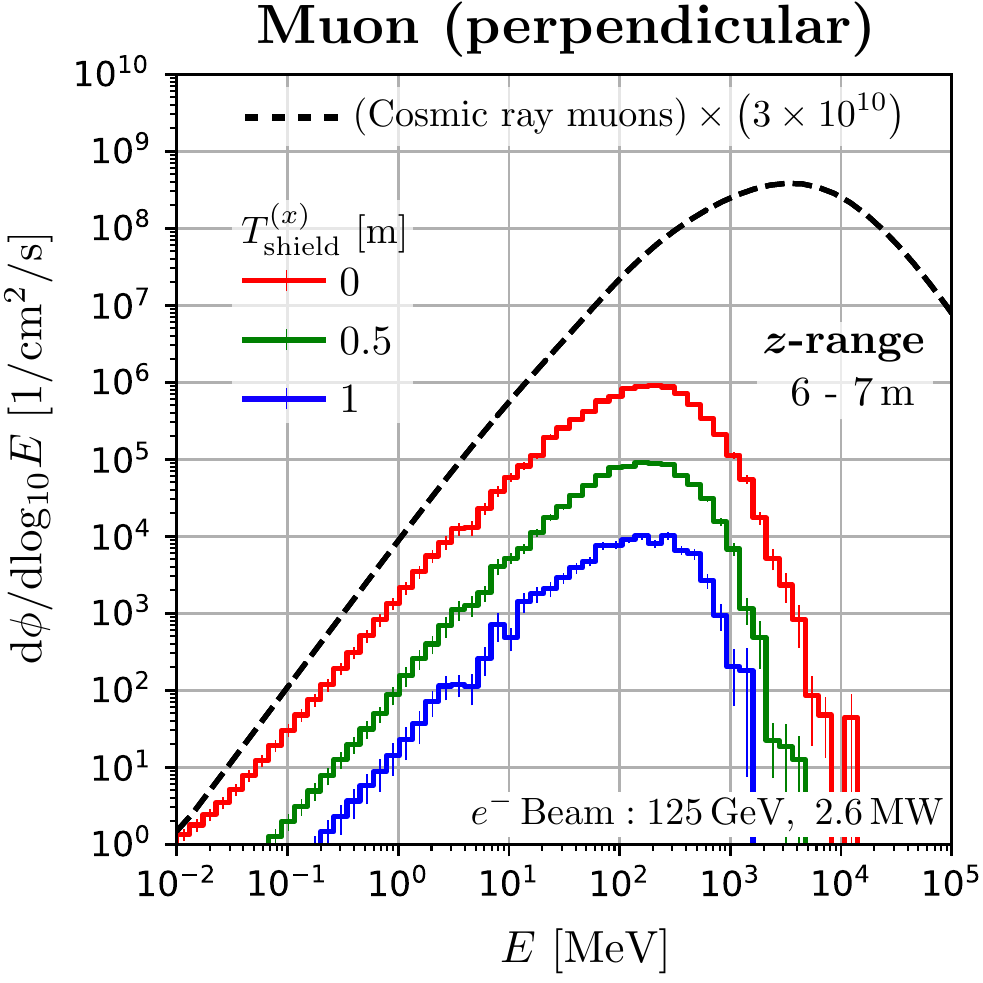}
\caption{The energy distribution of neutrons and muons at $z=6\text{ - }7$~m in the irradiation space of Geometry-2, where the dashed curves are the secondary cosmic ray at sea level, scaled by $3\times 10^{10}$.}
\label{side_E}
\end{center}
\end{figure}

We found that the neutrons in the perpendicular space have a spectrum quite similar to the secondary cosmic neutrons. Figure~\ref{side_E} shows the energy distribution of neutrons and muons at $z=6\text{ - }7$~m, where the dashed curves are the secondary cosmic rays at sea level multiplied by $3\times 10^{10}$, calculated with the {\tt EXPACS} code~\cite{sato2006analytical, sato2008development, sato2015analytical}. 
We can obtain the neutron spectra that perfectly agree with the secondary cosmic neutrons spectrum up to 2~GeV in the perpendicular space. Higher energy neutrons can also be obtained, although the spectra are different. These characteristics of the ILC cannot be obtained with other neutron irradiation facilities using low-energy protons.
The behavior of the high-energy tail of the neutron spectra depends slightly on the $z$-range of the irradiation space. 
We found that the ILC spectra are particularly consistent with that of cosmic-ray neutrons at $z=6\text{ - }7$~m.
This $z$-range is located at about 30 degrees to the beam axis from the region of maximum neutron production\footnote{$z=2\text{ - }3$~m on the beam axis.} in the beam dump. 
It can be seen that the spectra do not change significantly when the thickness of the shielding is changed. Spectra at different $z$-ranges are shown in Appendix~\ref{app:many_spectra}. From these findings, this perpendicular space is suitable for studying the soft errors induced by cosmic ray neutrons.

For the distributions of muons, the peak of the spectrum results from the combined effect of absorption at low energies and production spectrum at high energies. The peak position is of the order of total energy loss in the beam dump and the shield, which is a few 100 MeV in the perpendicular space. Therefore, it differs from the peak position of the secondary cosmic ray muons. However, the discrepancy can be improved in the region downstream of the beam dump, as described in the next section.

This agreement of the neutron spectra is qualitatively understandable. Secondary cosmic neutrons are induced by the primary cosmic protons. In contrast, in the case of electron incidence on beam dumps, neutrons are induced by bremsstrahlung photons. The energy dependence of the primary cosmic proton spectrum is $d\phi/dE \propto 1/E^p$, with $p$ ranging from 1 to 2.7 above 1~GeV, close to 2. On the other hand, the spectrum of bremsstrahlung in thick targets is $1/E^2$, and they show a similar distribution. Spectra of secondary hadrons produced by the interaction of the photons, protons and also other hadrons with matter typically have an energy dependence of $\sim 1/E$. Although the high energy proton emits high energy neutrons forward and breaks the $1/E$ structure, this is a minor effect in the perpendicular region. Moreover, secondary hadrons produce additional neutrons in the atmosphere or the water beam dump, which further average the neutron spectra. Thereby the difference in the produced neutron spectra due to the difference of the initial particles is lessened.

\subsection{Geometry-3: Flux in downstream space}
\label{sec:Geometry-3}

As in section~4.2, we perform the calculations using Geometry-3. The number of high-energy muons downstream of the beam dump differs between the electron and positron beams. In this section, we show the results for the electron beam dump. Details of the differences are given in Sec.~\ref{sec:muon}.

\begin{figure}[!t]
\begin{center}
\includegraphics[width=7.5cm, bb=0 0 326 283]{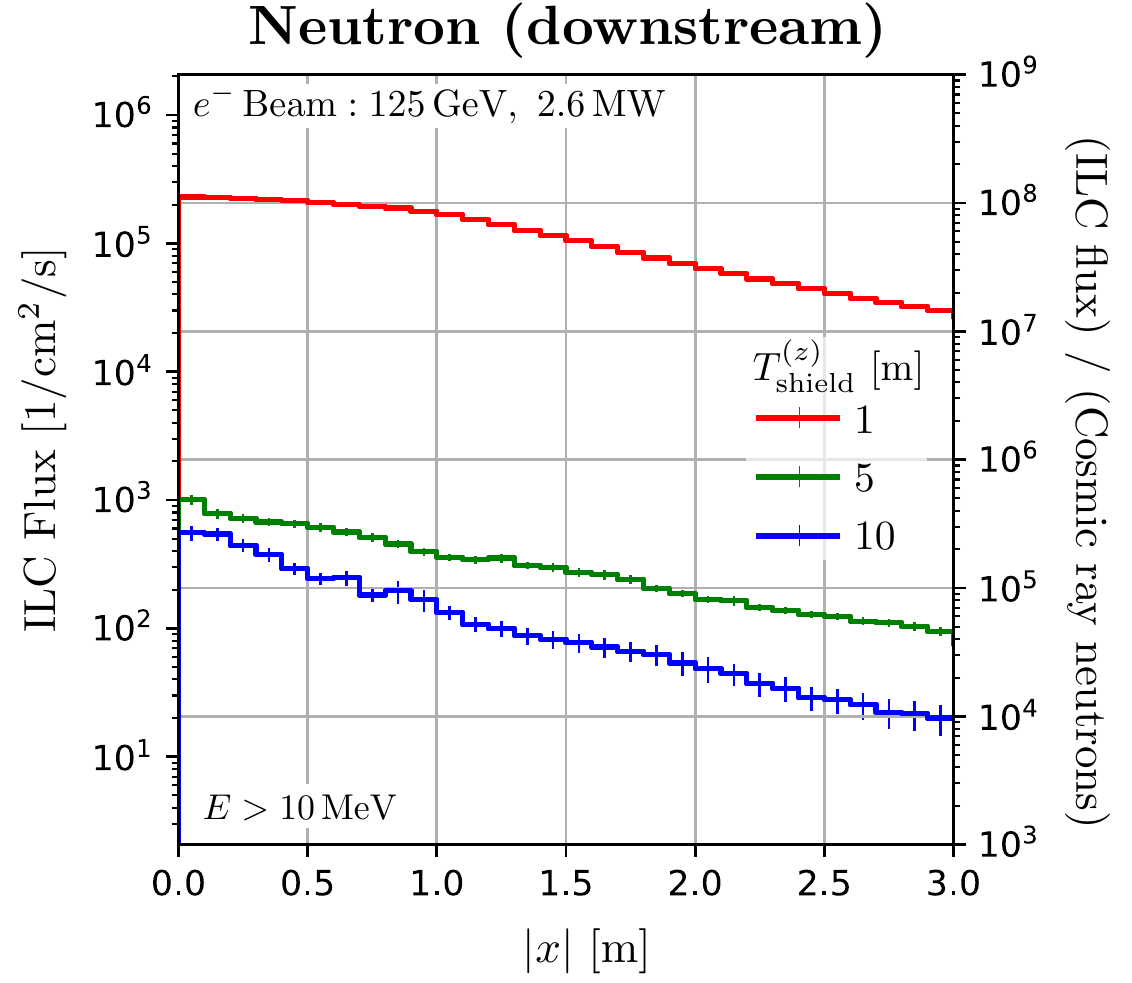}
\includegraphics[width=7.5cm, bb=0 0 326 283]{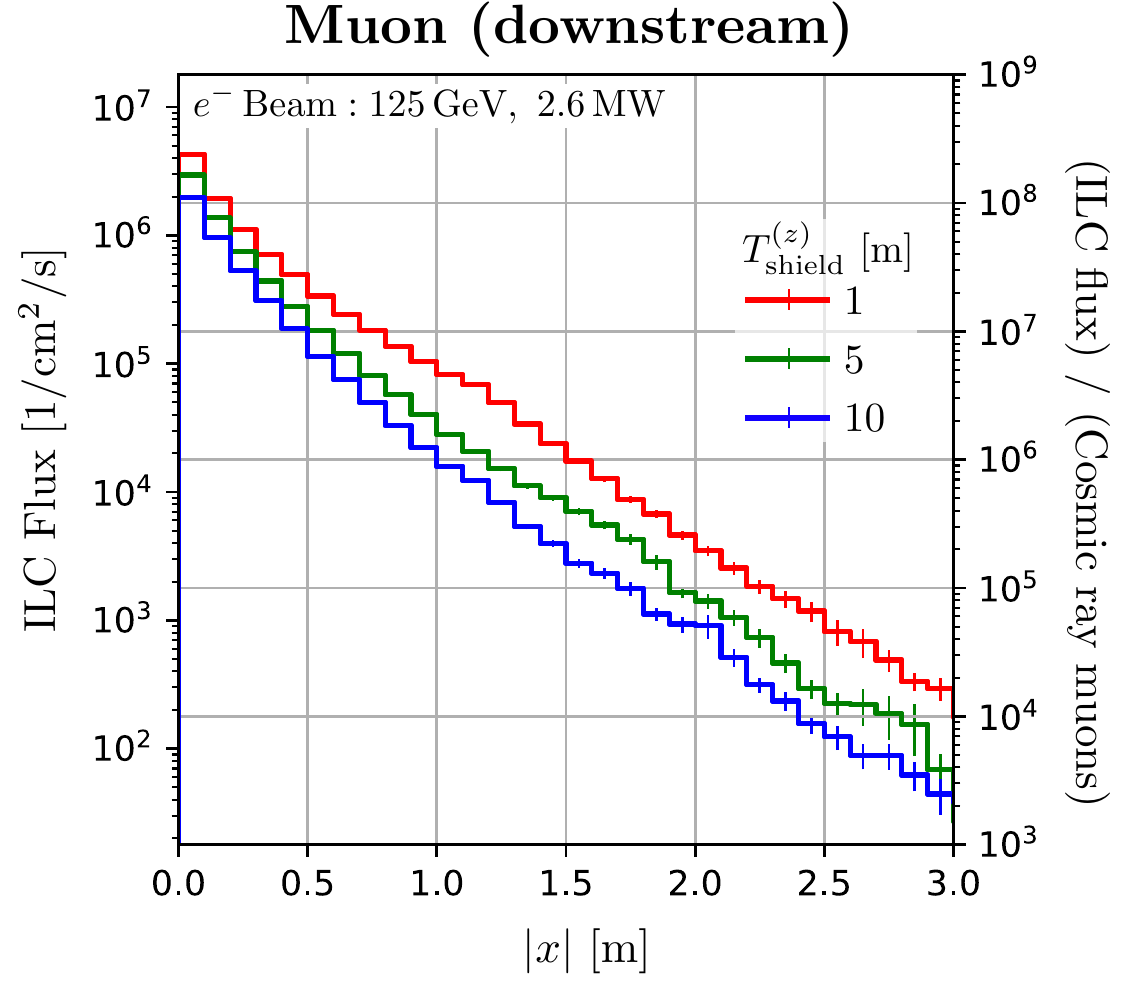}
\caption{The $x$ distribution of the flux in the irradiation space of Geometry-3, where $T_{\rm shield}^{(z)}=1, 5, $ and 10~m are chosen for the thickness of the concrete shield. The beam energy and power are 125~GeV and 2.6~MW.}
\label{back_x}
\end{center}
\end{figure}

Figure~\ref{back_x} shows the $x$-dependence of the flux in that irradiation space, where $T_{\rm shield}^{(z)}=1, 5, $ and 10~m are chosen for the thickness of the concrete shield. The flux is integrated over the width for $z$ and in the region $-50~{\rm cm}<y<50~{\rm cm}$. Especially, we can utilize high-intensity muon flux on the beam axis. The forward muons are more penetrating than neutrons, so the ratio between the fluxes of muons and neutrons can be optimized by changing the shielding thickness. The variation of muon flux intensity is within an order of magnitude in a region within 1\,m$^2$ centered on the beam axis, enabling large-area muon irradiation. If an irradiation sample larger than 1~m is preferred to be uniformly irradiated with muons, it may be necessary to rotate the target or vibrate it in the $x$ direction.

As can be seen from Figures~\ref{back_x} (and \ref{z}), the neutron attenuation per distance becomes more gradual in the region where the shielding thickness exceeds 5~m. This is because the neutrons in this region are induced by muon interaction with the nucleus.

\begin{figure}[!t]
\begin{center}
\includegraphics[width=7.0cm, bb=0 0 283 283]{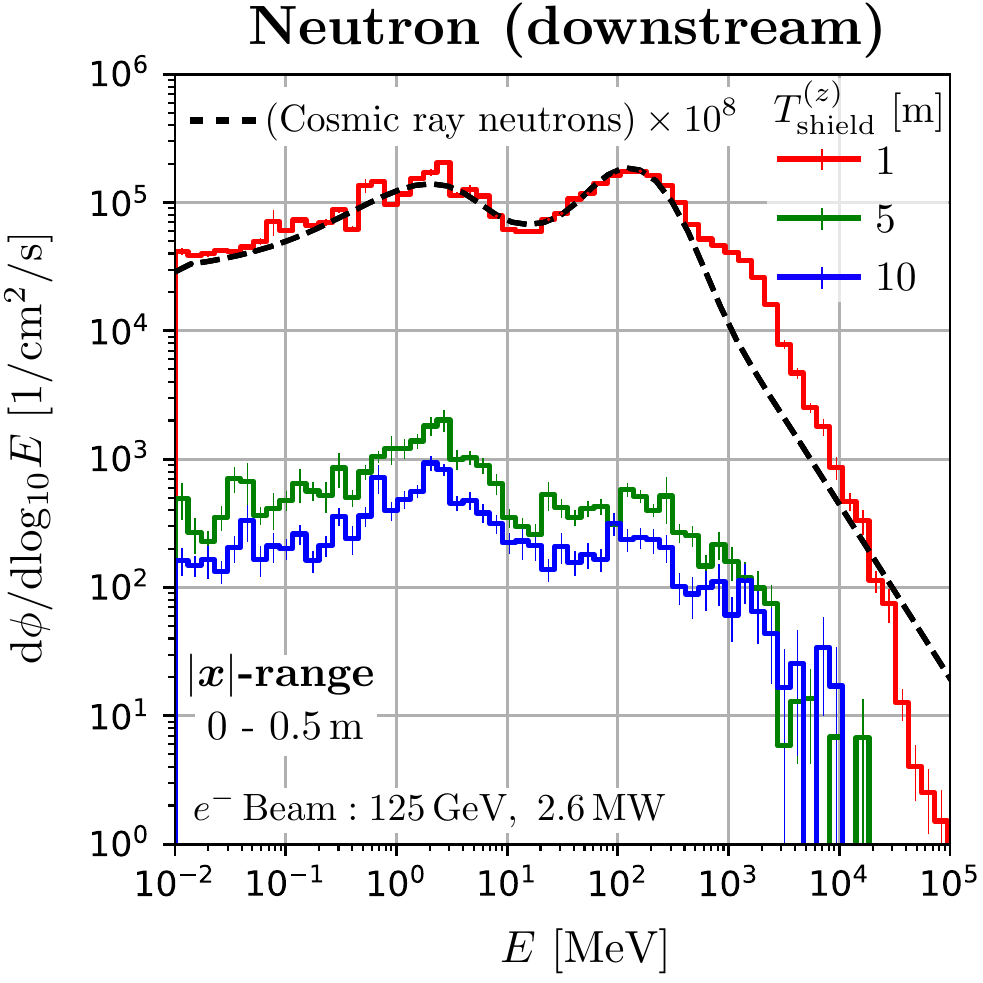}~~~
\includegraphics[width=7.0cm, bb=0 0 283 283]{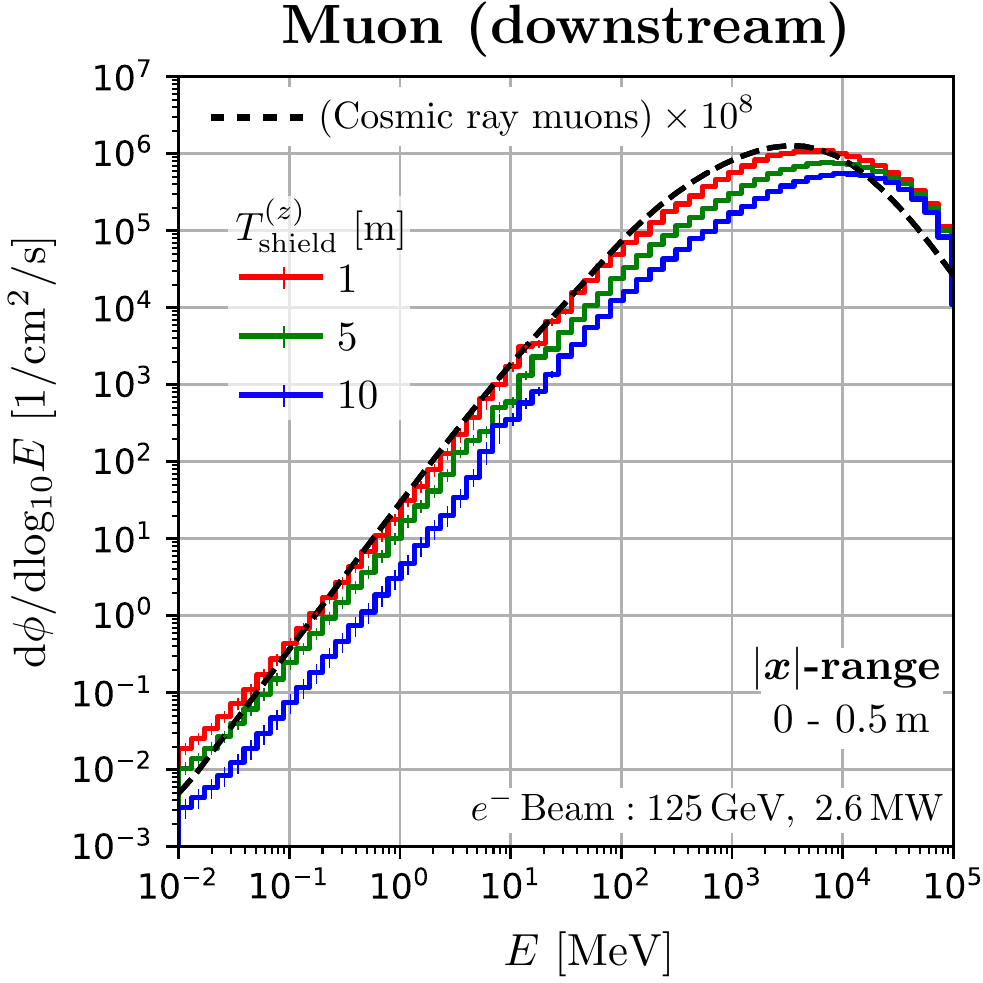}
\caption{The energy distribution of neutrons and muons at $|x|=0\text{ - }0.5$~m in the irradiation space of Geometry-3, where the dashed curves are the secondary cosmic ray at sea level, scaled by $10^{8}$.}
\label{back_E}
\end{center}
\end{figure}

We found that the muons and neutrons in this space have a spectrum quite similar to secondary cosmic rays. Figure~\ref{back_E} shows the energy distribution at $|x|=0\text{ - }0.5$~m in the space, where the dashed curves are the secondary cosmic ray at sea level in Tokyo, scaled by $10^8$. The neutron spectrum of the ILC is perfectly consistent with the secondary cosmic ray spectrum up to 400 MeV. Sufficient quantities of neutrons up to 40 GeV are available. Also, the high energy beam of the ILC provides sufficient high energy muons. The peak positions of the ILC muon spectra are close to that of the secondary cosmic ray muons. This is because the amount of material in the water beam dump is comparable to the amount of material in the atmosphere piled up on the ground. The energy at this peak position is close to the total energy that muons lose in the material. Therefore, this space is suitable for studying the effects of soft errors due to cosmic ray muons and neutrons. We want to emphasize that the ILC beam dump has the potential to be the only irradiation facility capable of providing high intensity, large area, and atmospheric-like muon flux.

\subsection{Beam Energy dependence}

\begin{figure}[!t]
\begin{center}
\includegraphics[width=7.5cm, bb=0 0 326 283]{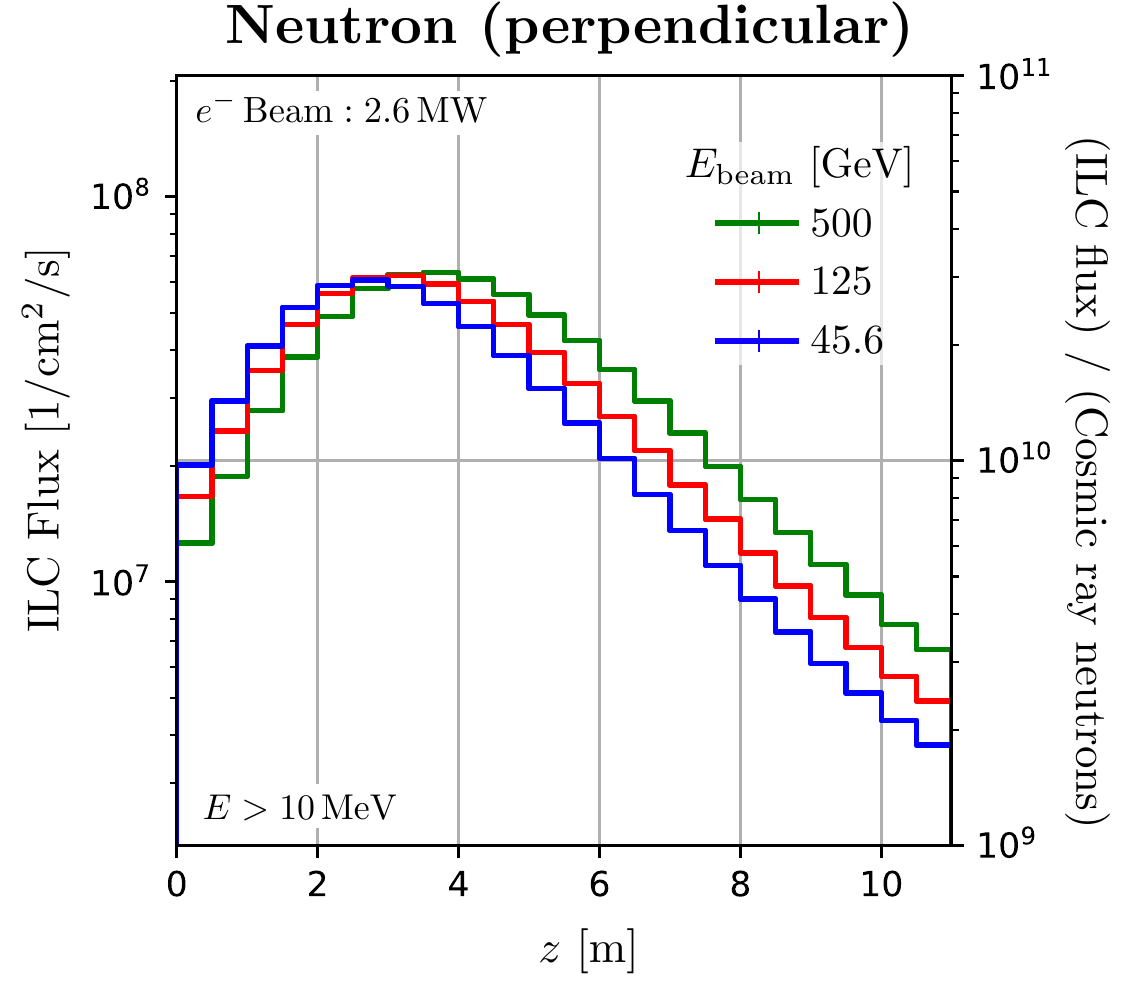}
\includegraphics[width=7.5cm, bb=0 0 326 283]{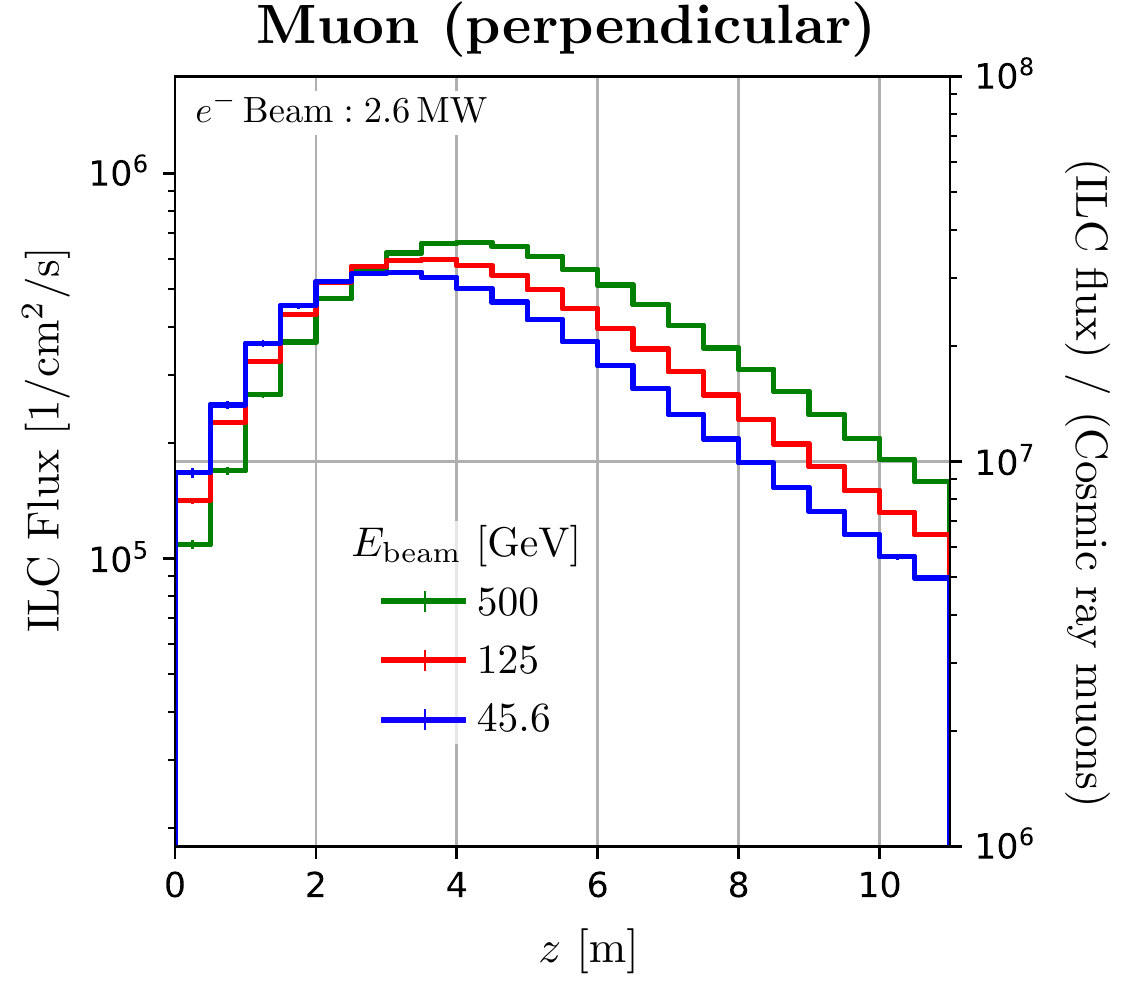}
\caption{The $z$ distribution of the flux in the irradiation space of Geometry-2, as in Figure~\ref{side_z}. The results are normalised to 2.6~MW beam power. The concrete thickness of $T_{\rm shield}^{(x)} = 0$~m is chosen.}
\label{Edep_side_z}
\end{center}
\end{figure}

\begin{figure}[!t]
\begin{center}
\includegraphics[width=7.0cm, bb=0 0 283 283]{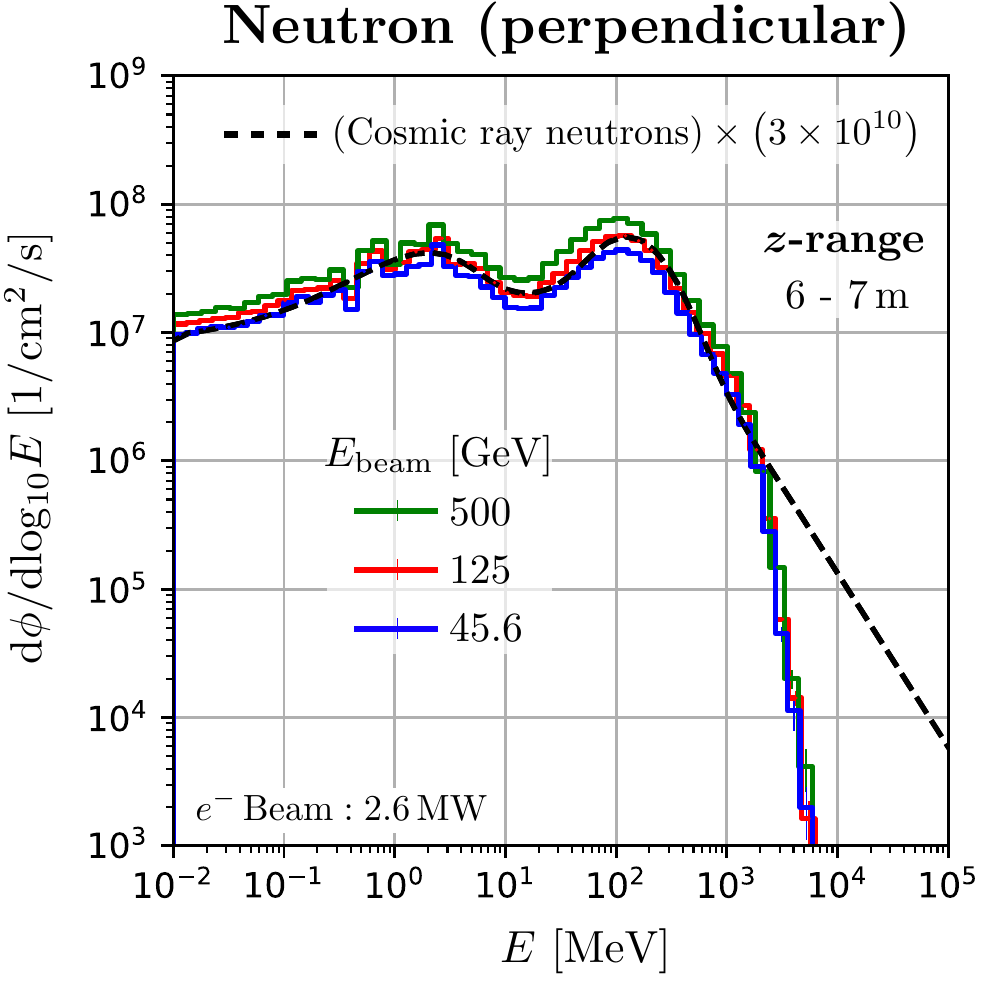}~~~
\includegraphics[width=7.0cm, bb=0 0 283 283]{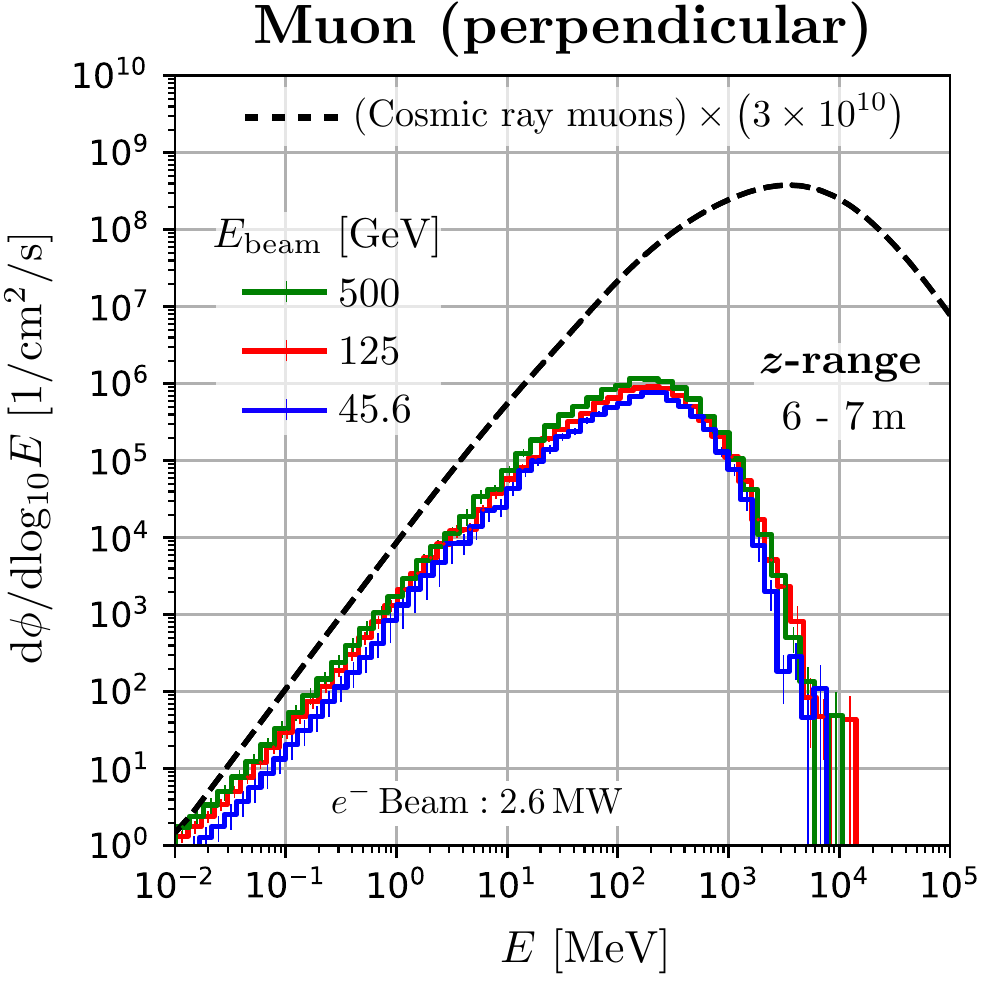}
\caption{The energy distribution of the fluxes for $E_{\rm beam}=500, 125, 45.6$~GeV in the irradiation space of Geometry-2, as in Figure~\ref{side_z}. The results are normalised to 2.6~MW beam power, and the concrete thickness of $T_{\rm shield}^{(x)} = 0$~m is chosen.}
\label{Edep_side_E}
\end{center}
\end{figure}

Several operations with different beam energies are planned at the ILC, as summarised in Table~\ref{table:ILC}. Neutron and muon fluxes for different beam energies are reported in this sub-section.

As the beam energy increases, the peak position of the flux slightly shifts downstream. Figure~\ref{Edep_side_z} is the $z$ distribution of the flux in the irradiation space of Geometry-2, as in Figure~\ref{side_z}. The results are normalised to 2.6~MW beam power, and a concrete thickness of $T_{\rm shield}^{(x)} = 0$~m is chosen. The peak position of the shower profiles in the beam direction is shifted by $X_0 \log(E_{\rm beam,2}/E_{\rm beam,1})$ for different beam energies $E_{\rm beam,1}$ and $E_{\rm beam,2}$~\cite{ParticleDataGroup:2020ssz}, where $X_0$ is the water radiation length.

The energy distribution of the flux in the perpendicular space is almost independent of the beam energy when normalised by beam power. Figure~\ref{Edep_side_E} shows the energy distribution for each beam energy in Geometry-2 at $z=6\text{ - }7$~m. Photons induce neutrons and muons in the perpendicular region with energies sufficiently lower than the beam energy. The number of low-energy photons is simply proportional to the beam power, so only the beam power is an important parameter.

\begin{figure}[!t]
\begin{center}
\includegraphics[width=7.0cm, bb=0 0 283 283]{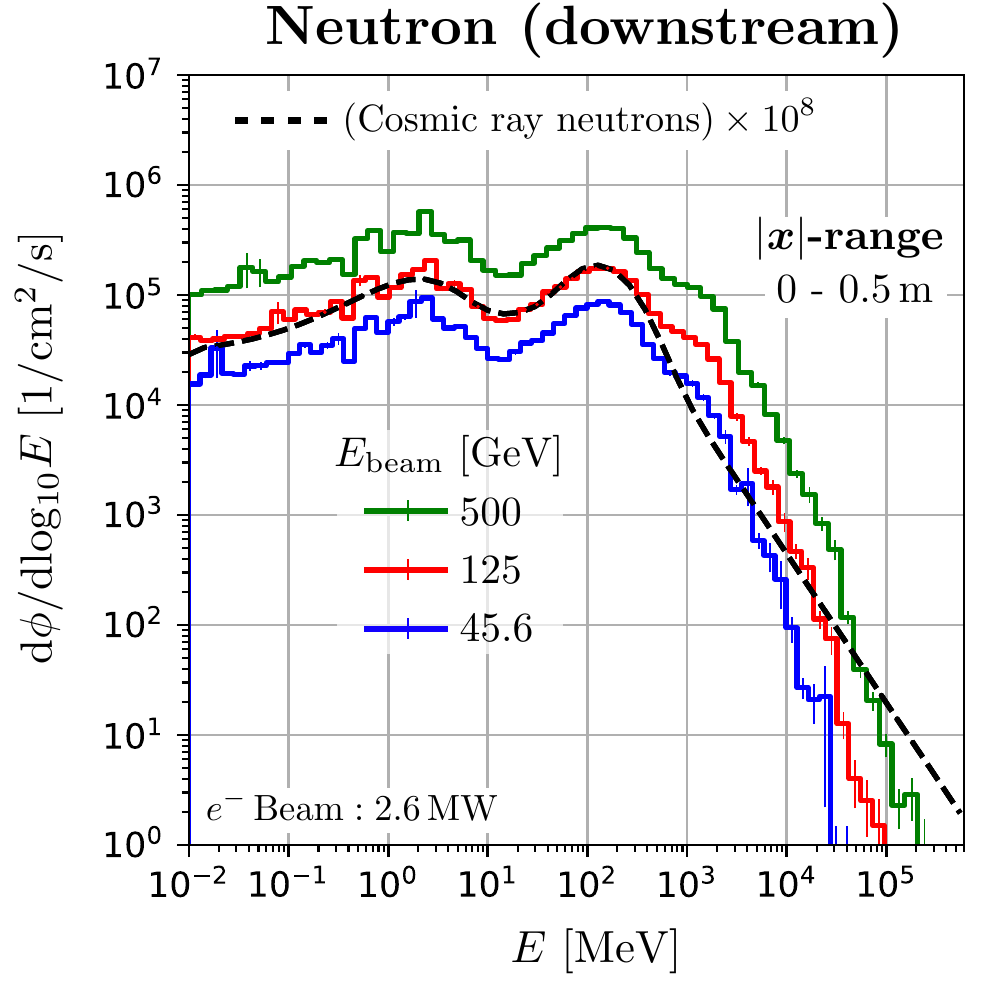}~~~
\includegraphics[width=7.0cm, bb=0 0 283 283]{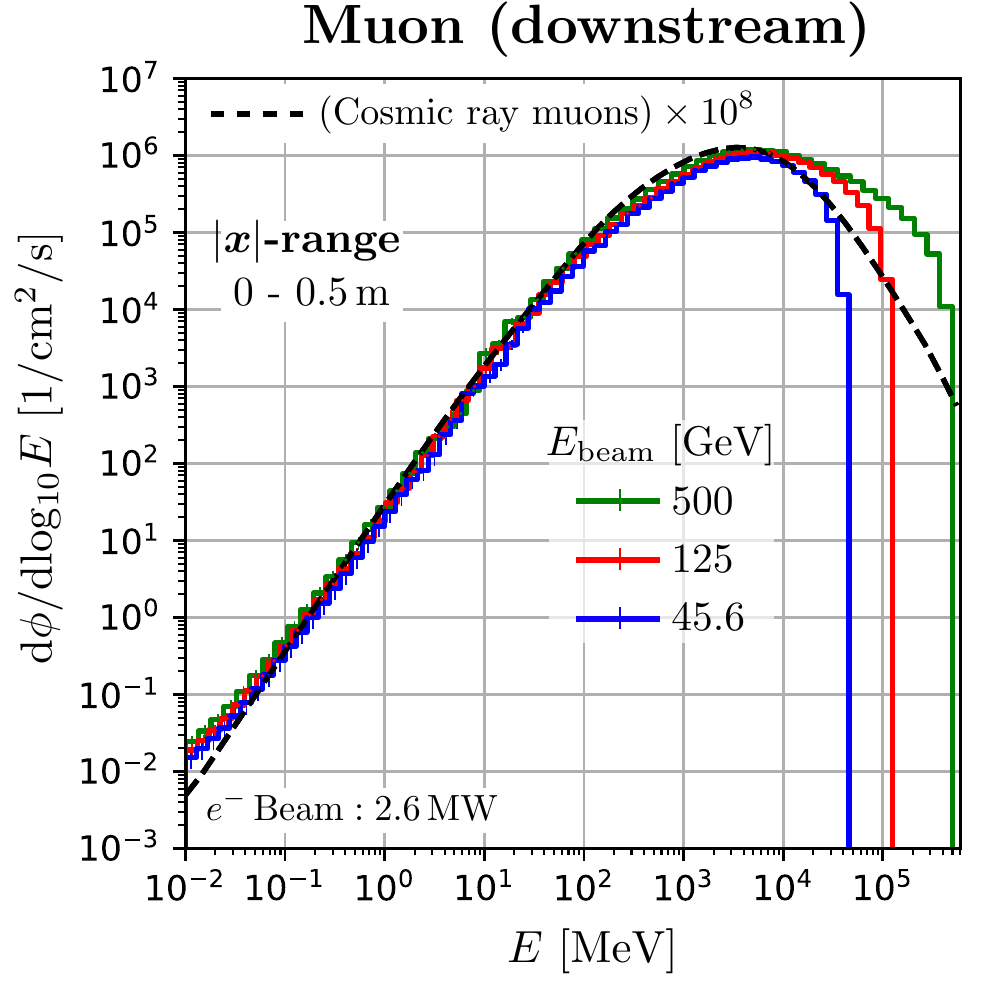}
\caption{The same plot as Figure~\ref{Edep_side_E} but for Geometry-3 and the concrete thickness of $T_{\rm shield}^{(z)} = 1$~m is chosen.}
\label{Edep_back_E}
\end{center}
\end{figure}

Figure~\ref{Edep_back_E} shows the energy distribution of the flux for each beam energy in the irradiation field of Geometry-3. All the results are normalised by 2.6~MW beam power. Downstream of the beam dump, it can be seen that the spectra of produced particles are similar to those of secondary cosmic rays for any ILC beam energy. The higher the beam energy, the higher the neutron flux, even when the beam power is normalised. This is because the multiplicity of neutrons produced in photonuclear reactions is greater at higher photon energies.
On the other hand, muons are induced by pair production, and the multiplicity of the muon is two. Therefore, the muon flux is approximately proportional to the beam power. We can also see that the high-energy tail of the muon spectrum extends up to the beam energy.

\subsection{Muon flux: e$^+$ vs e$^-$ beam dump}
\label{sec:muon}

In this section, we compare muon fluxes downstream of the electron beam dump and the positron beam dump. The differences are essential in experiments for particle physics using the beam dumps. There are two main processes for producing muons from positron beams or electromagnetic showers:
\begin{enumerate}
\renewcommand{\labelenumi}{(\arabic{enumi})}
\item Muon pair production induced by real photons $(\gamma \, {\rm N} \to \mu^+ \mu^- X)$;
\item Muon pair production induced by positrons $(e^+ e^-_{\rm atomic} \to \mu^+ \mu^-)$,
\end{enumerate}
where N denotes the nucleus. Process-(2) is caused by the pair annihilation of positrons above 43.7~GeV with atomic electrons in matter, which needs to be considered when using high-energy positron beams.

\begin{figure}[!t]
\begin{center}
\includegraphics[width=7.0cm, bb=0 0 314 297]{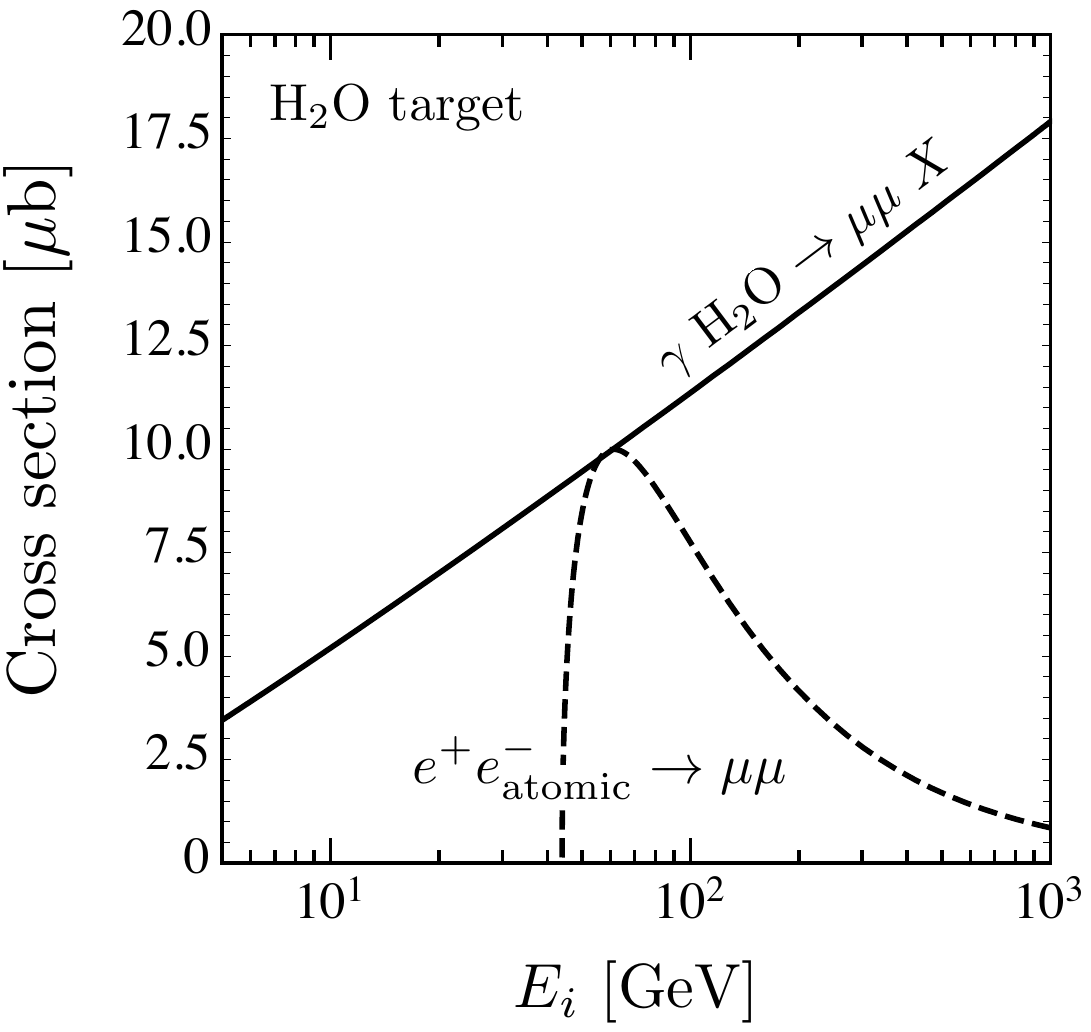}~~~
\includegraphics[width=7.0cm, bb=0 0 312 303]{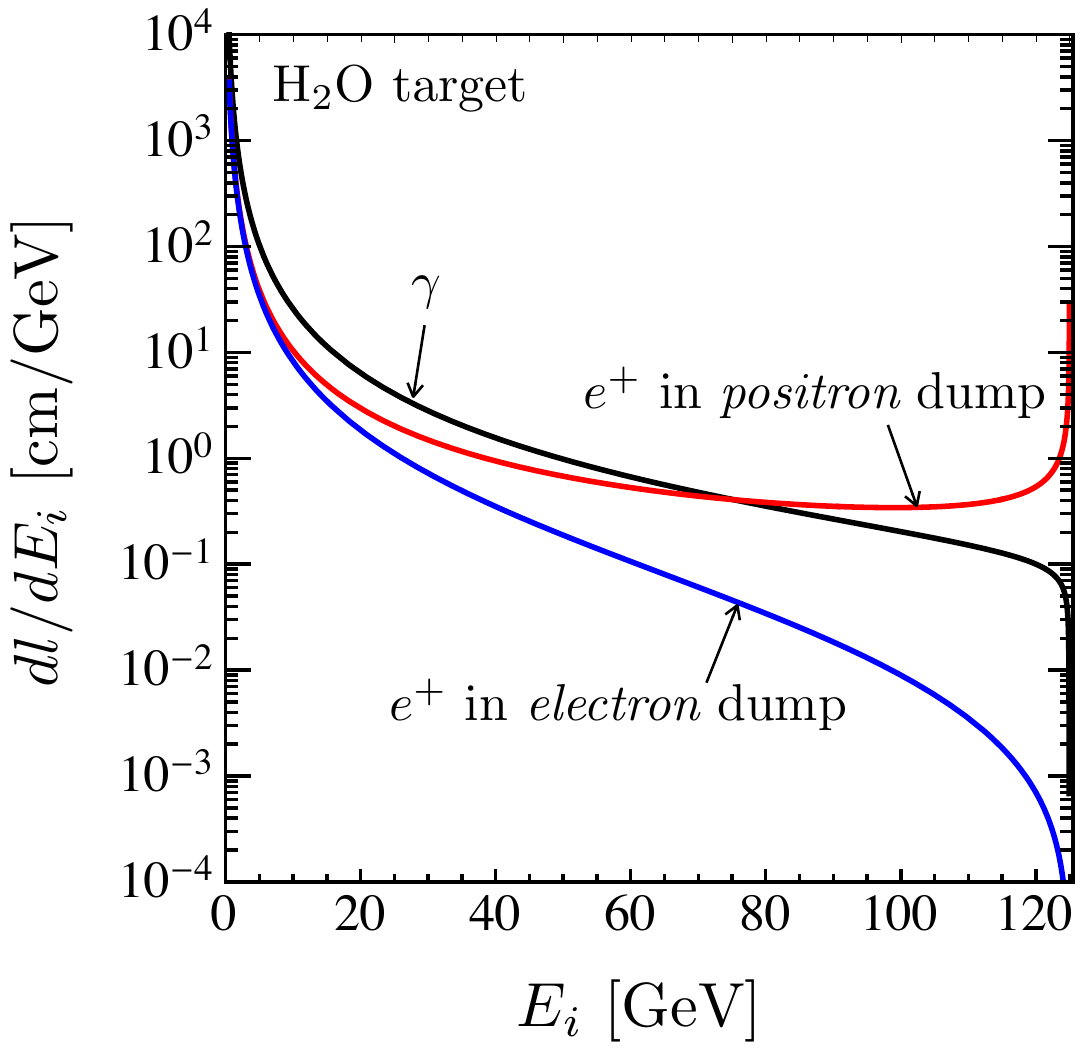}
\caption{
{\bf [Left]} The muon production cross section for one water molecule for Process (1) $\gamma \, {\rm H_2O} \to \mu^+ \mu^- X$ and (2) $e^+ e^-_{\rm atomic} \to \mu^+ \mu^-$. 
{\bf [Right]} The track lengths of photons and positrons in the electron and positron beam dump for $E_{\rm beam}=125$~GeV. 
}
\label{2process}
\end{center}
\end{figure}

The muon production cross section for one water molecule for each process is represented in Figure~\ref{2process} (Left panel). The energies $E_i$ are those of the initial state photon and positron for processes (1) and (2). The corresponding cross sections per water molecule are given approximately by~\cite{Tsai:1973py}
\begin{align}
\sigma_{\rm H_2O}^{(1)} 
&\simeq (2Z_{\rm H}^2+Z_{\rm O}^2) \hat{\sigma}_{\gamma\to\mu\mu}^{\rm (coherent)},\\
\sigma_{\rm H_2O}^{(2)} 
&\simeq (2Z_{\rm H}  +Z_{\rm O}  ) \sigma_{ee\to\mu\mu},
\end{align}
where $\hat{\sigma}_{\gamma\to\mu\mu}^{\rm (coherent)}$ is the coherent interaction between a real photon and a nucleus, neglecting the atomic number dependent term, and $Z_{\rm H}$ and $Z_{\rm O}$ are the atomic numbers of hydrogen and oxygen. From the above results, cross sections for any atom or molecule can be approximatively evaluated.

\begin{figure}[!t]
\begin{center}
\includegraphics[width=14.0cm, bb=0 0 1004 412]{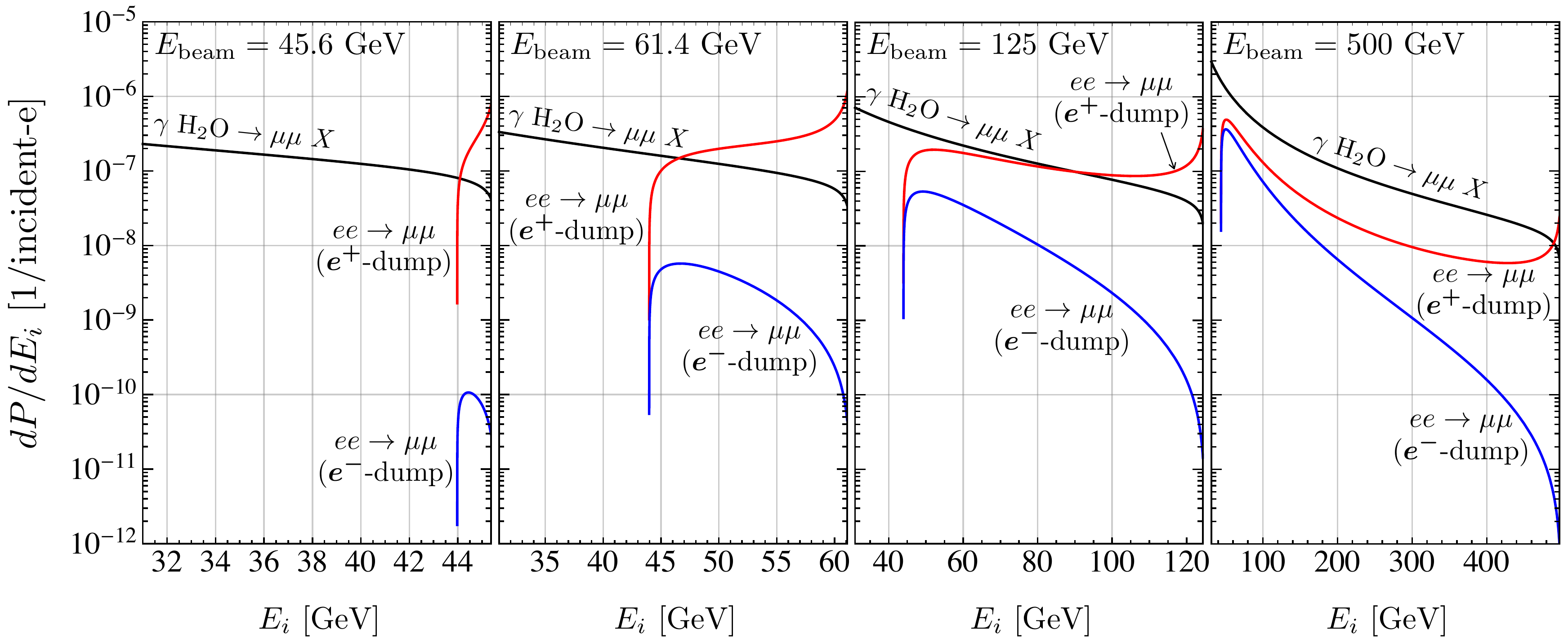}
\caption{The energy distribution of the number of muons generated per injection into the beam dump for beam energies of 45.6, 61.4, 125 and 500~GeV.}
\label{2process_dist}
\end{center}
\end{figure}

The energy distribution of the number of muons generated per injection into the beam dump is expressed by
\begin{align}
\frac{dN_{\mu}^{(1,2)}}{dE_i} = n_{\rm H_2O} \frac{dl}{dE_i} \sigma_{\rm H_2O}^{(1,2)}
\end{align}
$n_{\rm H_2O}$ is the number density of water mole, $dl/dE_i$ is the track length of particle $i$ in the beam dump. The track lengths of photons and positrons in the electron and positron beam dump~\cite{Asai:2021ehn} are given in Figure~\ref{2process} (Right panel). Figure~\ref{2process_dist} shows the energy distribution of the calculated number of muons. The results are for beam energies of 45.6, 61.4, 125 and 500~GeV, where 61.4~GeV is the energy that gives the largest cross section for Process-(2). The black line represents the contribution from Process-(1), and the results are equal for electron or positron beam dumps. The red and blue lines represent the contribution from Process-(2) in the electron or positron beam dump. Process-(2) contributes significantly to the high energy region in the positron beam dump, especially when the beam energy is around 100~GeV.

\begin{figure}[!t]
\begin{center}
\includegraphics[width=7.0cm, bb=0 0 400 310]{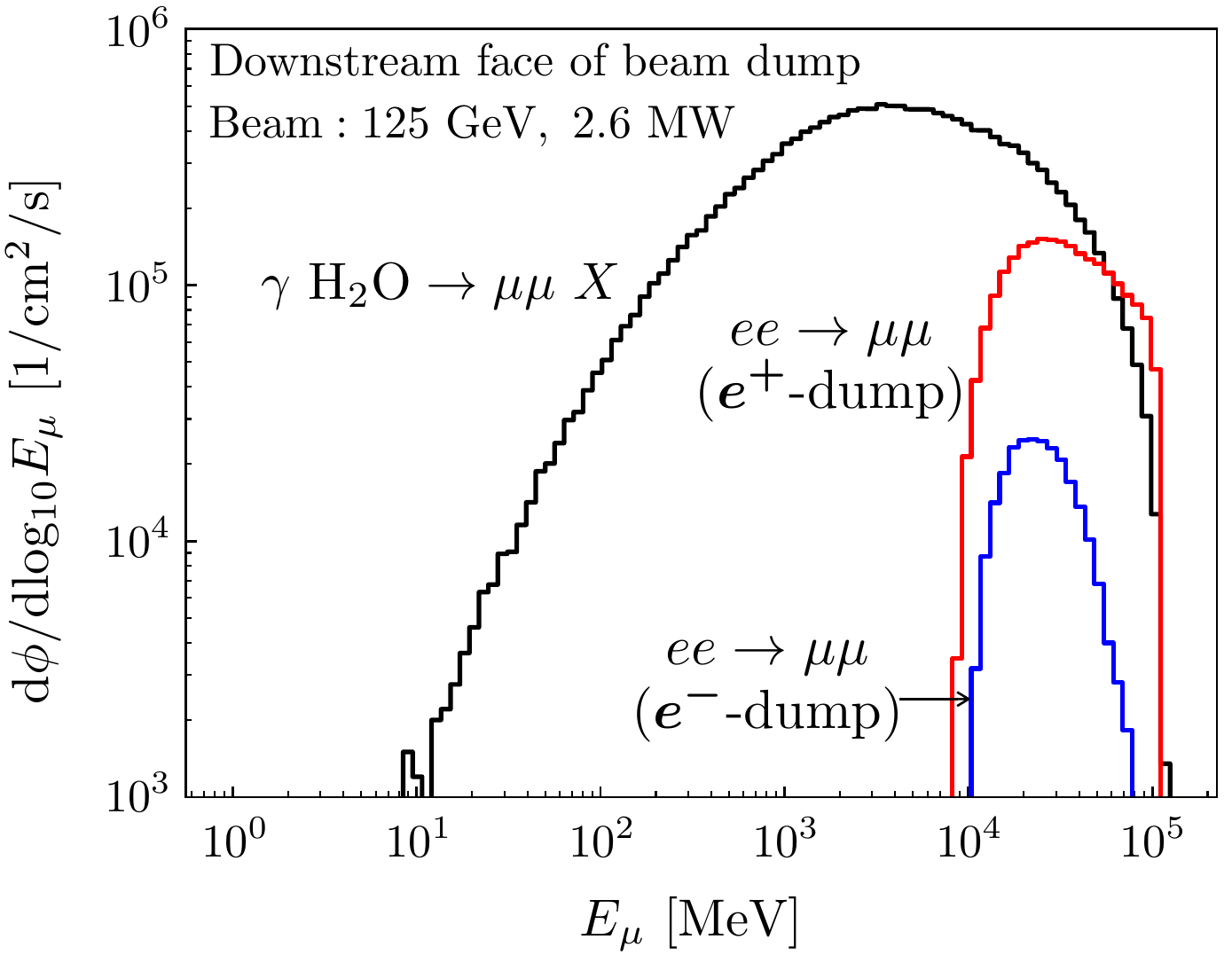}
\caption{The kinetic energy distribution of muons penetrating the downstream face of the beam dump cylinder $(z=11~{\rm m}, r<0.9~{\rm m})$.}
\label{2process_E}
\end{center}
\end{figure}

\begin{figure}[!t]
\begin{center}
\includegraphics[width=13.0cm, bb=0 0 943 590]{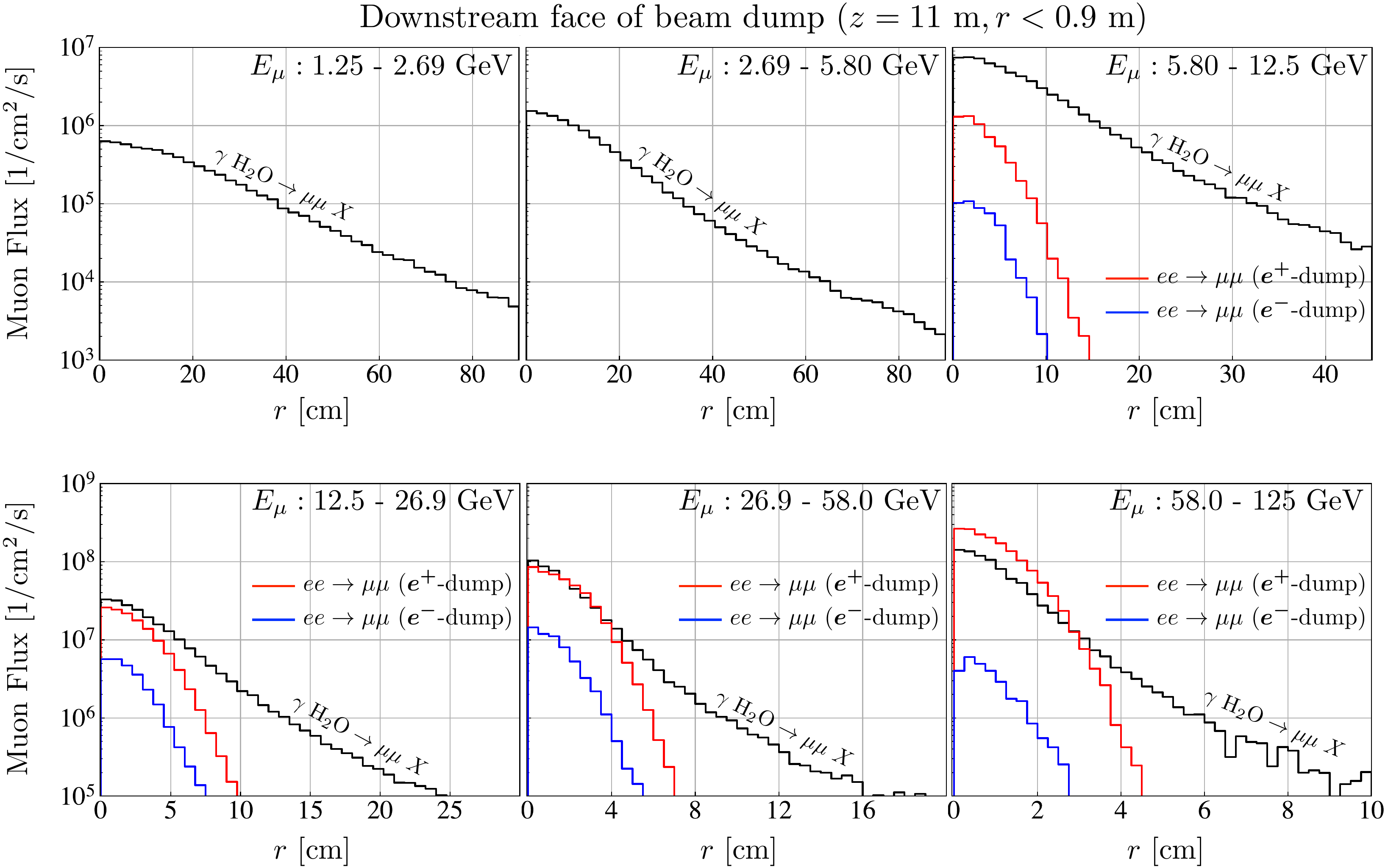}
\caption{Muon radial distribution around the beam axis for different energy ranges.}
\label{2process_r}
\end{center}
\end{figure}

We investigate the detailed energy distribution and spatial extent of the muons penetrating the downstream face of the beam dump cylinder $(z=11~{\rm m}, r<0.9~{\rm m})$, where the spatial extent is expressed as the distance of the muon from the beam axis, $r=\sqrt{x^2+y^2}$. This information is essential for particle physics experiments. Figure~\ref{2process_E} shows the kinetic energy distribution of the muon flux. It can be seen that Process-(2) is dominant at high energies in positron beam dumps. Figure~\ref{2process_r} shows the muon radial distribution around the beam axis for different energy ranges. The muons are concentrated in the forward region over a wide energy range. In particular, muons produced in Process-(2) are more likely to be scattered forward than those in Process-(1).

\clearpage

\section{Summary}

We evaluated the neutron and muon fluxes produced in the ILC beam dumps by Monte Carlo simulations and considered their potential use in irradiation fields for soft error studies. The basic design of the current ILC beam dump is an 11~m long cylindrical tank filled with water as an absorber. 
This length of water is equivalent to 100~km of atmospheric material piled up on the ground, so the ILC beam dump can be expected to act as a generator of secondary particles with atmospheric-like spectra.

In the beam dump facilities, irradiation spaces can be located perpendicular to the beam axis or downstream of the beam dump, and the fluxes in these spaces were evaluated. The respective simulation geometries have been represented in Figure~\ref{Geometry}.

In the region perpendicular to the beam axis, a neutron intensity up to $10^{11}$ times higher than the secondary cosmic ray is available. The neutron spectrum produced in the ILC beam dump is similar to that of the secondary cosmic neutrons, agreeing perfectly up to a few GeV. The variation of neutron flux intensity is within an order of magnitude over 10~m length, which enables uniform large-area neutron irradiation. Therefore, this space is suitable for studying the soft errors induced by cosmic ray neutrons and can accommodate the irradiation of large volume samples.

In the downstream space, muons and neutrons with intensities $10^8$ times higher than secondary cosmic rays are available, and their spectra are similar to those of secondary cosmic rays. The variation of muon flux intensity is within an order of magnitude in a region of 1\,m$^2$ centered on the beam axis, enabling large-area muon irradiation. The forward muons are more penetrating than neutrons, so the number of neutrons can be adjusted by changing the shielding thickness while keeping the number of muons. This space has the potential to become a distinctive facility capable of irradiating high-intensity atmospheric muons in sizes on the order of 1\,m$^2$, which is not possible in existing irradiation facilities.

Assuming several ILC operations, we evaluated the fluxes at 45.6, 125 and 500~GeV beam energies. We found that at all beam energies, the ILC beam dumps produce a spectrum similar to the secondary cosmic ray. The flux is essentially proportional to the beam power, but due to the increase in neutron multiplicity per reaction, the higher the beam energy, the more neutrons are obtained.

We also discussed the difference in the energy distribution of muons produced in the electron and positron beam dump. Positrons above 43.7~GeV annihilate with atomic electrons in matter, producing muon pairs. We showed that downstream of the positron beam dump, the muons in the high-energy region are dominated by the positron-induced muons.

\subsection*{Acknowledgement}
We would like to thank Takashi Nakano, Masanori Hashimoto, Yoshihito Namito, and Hitoshi Murayama for helpful discussions and comments.

\appendix

\begin{figure}[!t]
\begin{center}
\includegraphics[width=7.0cm, bb=0 0 283 283]{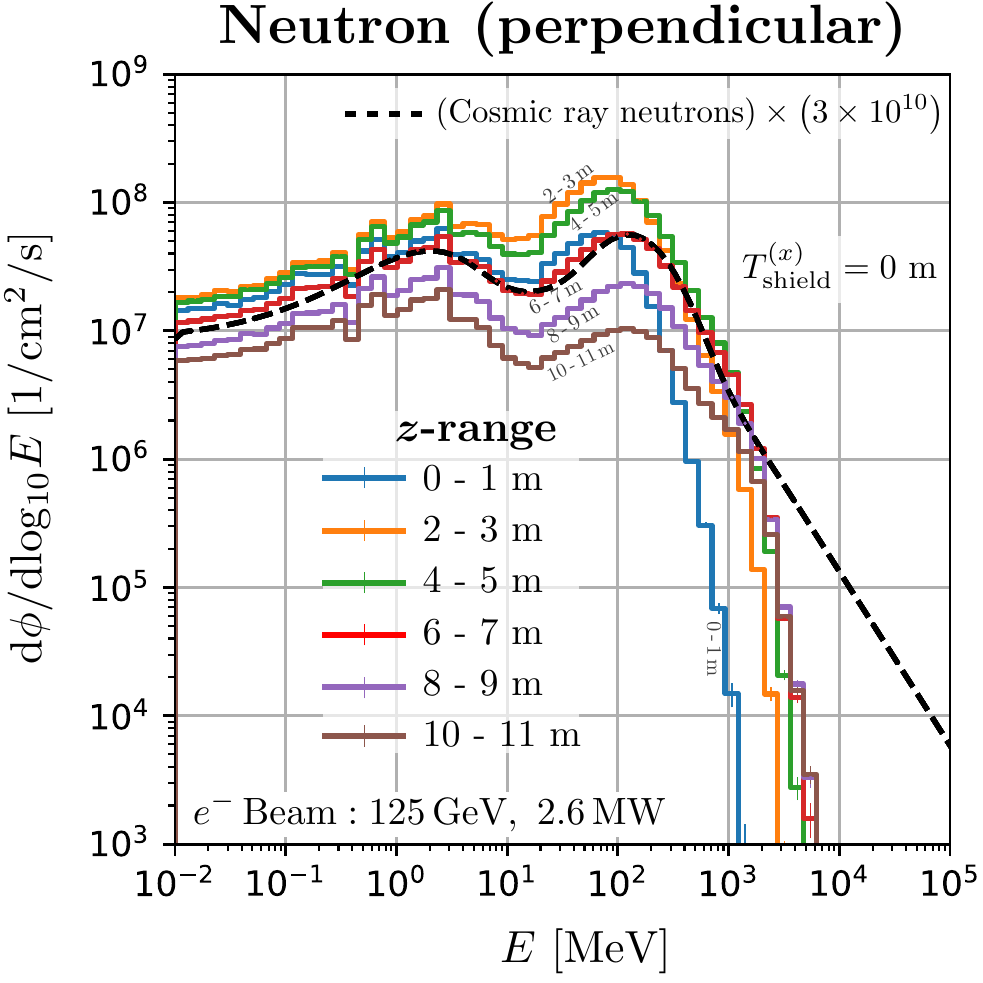}~~~
\includegraphics[width=7.0cm, bb=0 0 283 283]{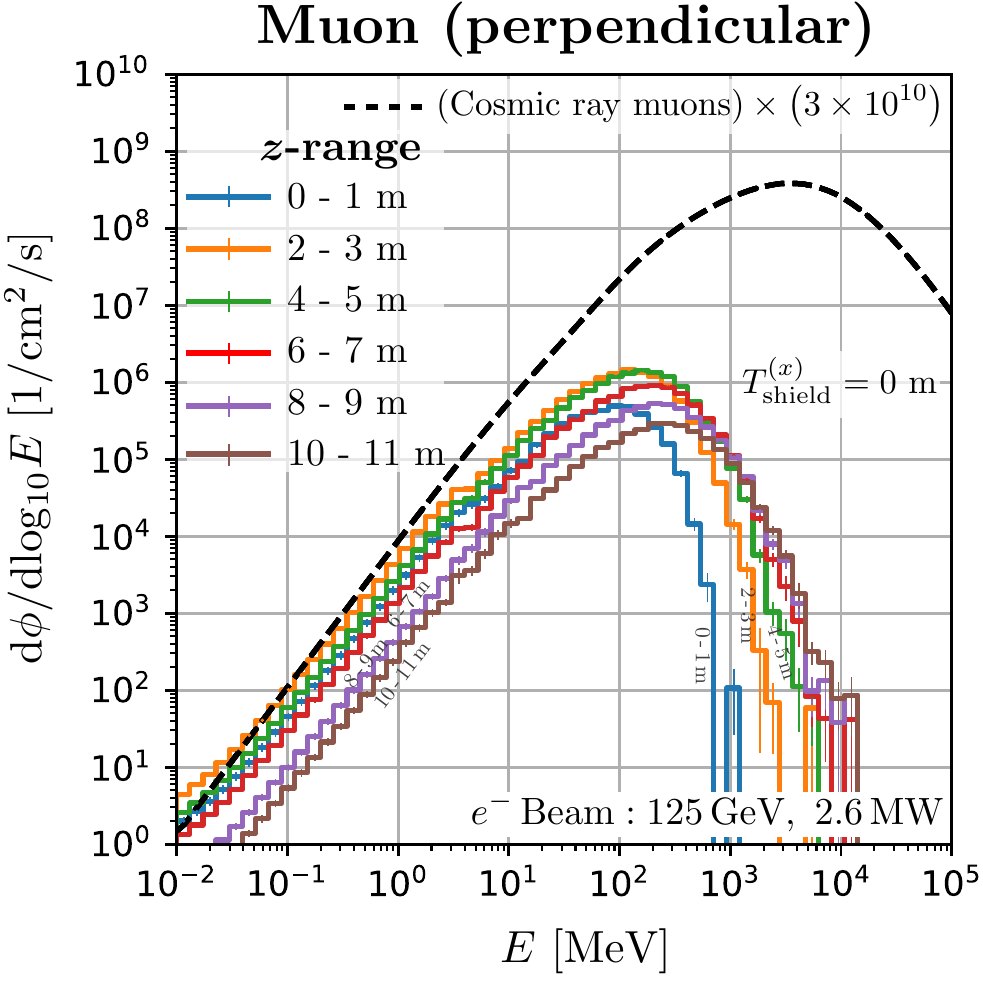}
\caption{Neutron and muon spectra for various z-ranges from $z=0$ to 11~m for Geometry-2 in Figure~\ref{Geometry}. The concrete thickness in $x$ direction is $T_{\rm shield}^{(x)}=0$~m.}
\label{side_E_more}
\end{center}
\end{figure}

\begin{figure}[!t]
\begin{center}
\includegraphics[width=7.0cm, bb=0 0 283 283]{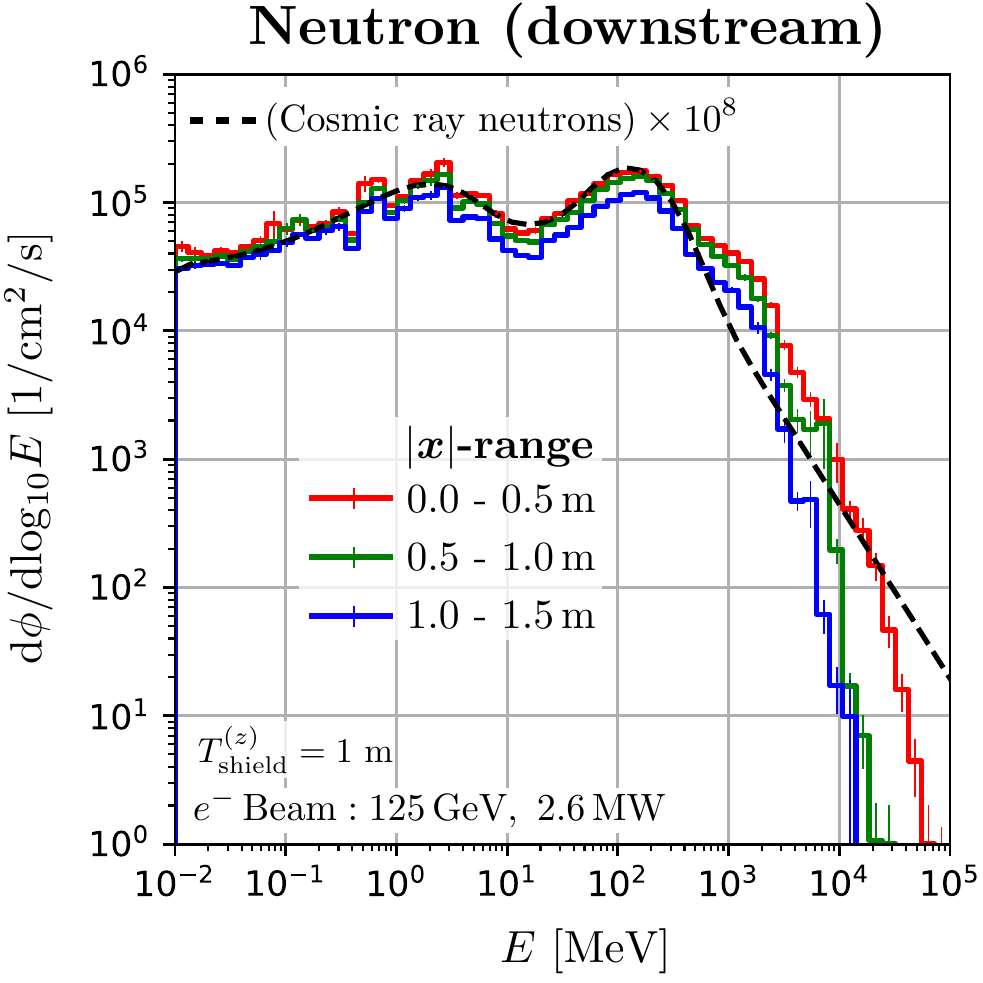}~~~
\includegraphics[width=7.0cm, bb=0 0 283 283]{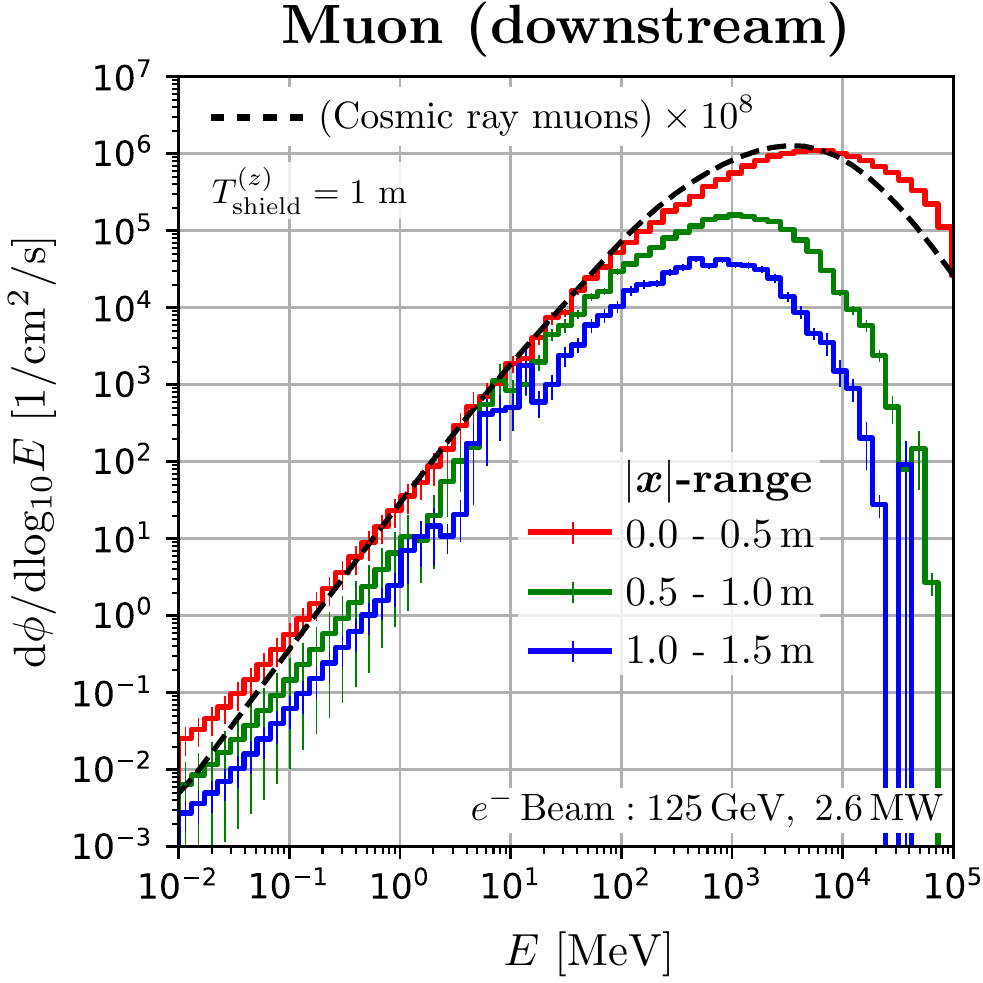}
\caption{Neutron and muon spectra for the three $|x|$-range for Geometry-3 in Figure~\ref{Geometry}. The concrete thickness in $z$ direction is $T_{\rm shield}^{(z)}=1$~m.}
\label{back_E_more}
\end{center}
\end{figure}

\section{Spectra at various regions around the beam dump}
\label{app:many_spectra}

In Sections \ref{sec:Geometry-2} and \ref{sec:Geometry-3}, the neutron and muon spectra for some regions around the beam dump are presented: in the case of Sections~\ref{sec:Geometry-2}, the results are for $6~{\rm m}<z<7~{\rm m}$ in Geometry-2 of Figure~\ref{Geometry}, and in the case of Section~\ref{sec:Geometry-3}, for $|x|<0.5~{\rm m}$ in Geometry-3. This section presents results for other $z$-ranges and $|x|$-ranges.

Figure~\ref{side_E_more} shows neutron and muon spectra for various $z$-ranges from $z=0$ to 11~m for Geometry-2 in Figure~\ref{Geometry}. The results for $T_{\rm shield}^{(x)}=0$~m are shown, and the normalization of the spectra decreases as the shield is thickened, as shown in Figure~\ref{side_E}. For neutrons, the spectrum for the $z=6~\text{-}~7$~m range is most consistent with the cosmic ray neutrons. The spectra for other ranges also have a shape similar to the cosmic spectrum. For all ranges, there is a peak around 100~MeV, and the number of neutrons above 100~MeV for that peak value increases as $z$ increases. This can be explained through kinematics. The amount of neutrons below 10~MeV also depends slightly on $z$. For muons, the spectra do not agree with the cosmic muons in the perpendicular region. As in the case of neutrons, the fraction of high-energy muons increases as $z$ increases.

Figure~\ref{back_E_more} shows the neutron and muon spectra for the three $|x|$-range for Geometry-3 in Figure~\ref{Geometry}. This is the result for $T_{\rm shield}^{(z)}=1~\text{m}$. For the muon spectra, the spectrum's shape is highly dependent on $|x|$. The muons in this region are due to pair production from bremsstrahlung. The bremsstrahlung is emitted in the direction of the beam axis, and the typical production angle of muons is $m_{\mu}/E_{\mu}$, where $m_{\mu}$ is the muon mass, and $E_{\mu}$ is the muon energy. Therefore, high-energy muons are distributed around the beam axis. With respect to the neutron spectra, the fraction of high-energy neutrons increases slightly as $|x|$ decreases.

\bibliographystyle{unsrt}
\bibliography{refs}

\end{document}